\documentclass[prd,preprint,tightenlines,floatfix,showpacs,preprintnumbers,nofootinbib,eqsecnum,superscriptaddress]{revtex4-1}

 \usepackage[dvips,final]{graphicx}
  \usepackage{amssymb}
   \usepackage{amsmath}
    \usepackage{amsfonts}
     \usepackage{epsfig}
      \usepackage{bm}

\def\Pom{{\bf I\!P}}

\def\lsim{\mathrel{\rlap{\lower4pt\hbox{\hskip1pt$\sim$}}
    \raise1pt\hbox{$<$}}}         
\def\gsim{\mathrel{\rlap{\lower4pt\hbox{\hskip1pt$\sim$}}
    \raise1pt\hbox{$>$}}}         

\begin{document}

\title{Inclusive and exclusive diffractive production of dilepton pairs
  \\  in proton-proton collisions at high energies}

\author{G.~Kubasiak}
\email{Gabriela.Slipek@ifj.edu.pl}
\affiliation{Institute of Nuclear Physics PAN, PL-31-342 Cracow,
Poland} 
\author{A.~Szczurek}
\email{Antoni.Szczurek@ifj.edu.pl}
\affiliation{Institute of Nuclear Physics PAN, PL-31-342 Cracow,
Poland} 
\affiliation{University of Rzesz\'ow, PL-35-959 Rzesz\'ow,
Poland}

\date{\today}

\begin{abstract}
We calculate for the first time cross sections for
single and central diffractive as well as exclusive
diffractive production of dilepton pairs in proton-proton collisions. 
Several differential distributions are shown. 
The inclusive diffractive processes are calculated
using diffractive parton distributions extracted from
the analysis of diffractive structure function and dijet
production at HERA.
We find that the inclusive single-diffractive Drell-Yan process is 
by about 2 orders of magnitude smaller than ordinary Drell-Yan process.
The central-diffractive processes are smaller by one order
of magnitude compared to single-diffractive ones.
We consider also exclusive production of dilepton pairs.
The exclusive photon-pomeron (pomeron-photon) process
constitutes a background to the QED photon-photon process
proposed to be used for controlling luminosity at LHC. 
Both processes are compared then in several differential 
distributions. We find a region of the phase space where 
the photon-pomeron or pomeron-photon contributions can be larger than 
the photon-photon one.
\end{abstract}

\pacs{12.40.-y, 13.60.-r, 13.85.Qk}

\maketitle

\section{Introduction}

The Drell-Yan production process is often used to extract
quark and antiquark distributions in the nucleon. Next to open heavy quark
production associated with heavy-flavour semileptonic decays it is 
one of the most important mechanisms for inclusive production of
leptons. At large lepton rapidity the Drell-Yan process may be
very sensitive to the gluon distributions at small-x \cite{small_x}.
Can the diffractive mechanisms contribute to this region too? 

The diffractive processes were intensively studied in $e p$ collisions
at HERA. A formalism has been developed how to calculate them
in terms of the diffractive structure functions.
The situation in proton-proton collisions is more complicated.
The  single production diffractive cross sections in proton-proton collisions
constitute usually less than 5 \% of standard inclusive cross sections.
They were calculated for $W$ and $Z$ boson \cite{diffractive_gauge_bosons},
 dijet \cite{diffractive_dijets}, 
open $c \bar c$ \cite{diffractive_open_charm}
 as well as for Higgs \cite{diffractive_Higgs}.
It was shown that a naive Regge factorization leads to a sizeable
overestimation of the cross section and additional absorption
mechanism must be included.
Central diffractive processes where calculated only for jet 
\cite{double_pomeron}, $c \bar c$ \cite{double_diff} and 
$Z^0$ \cite{CSS09}.
Here we wish to calculate their contribution for dilepton production.
In this context we will use diffractive parton distributions
found by the H1 Collaboration in the analysis of proton
diffractive structure function $F_2^{(D)}$ as well as 
dijet production in DIS \cite{H1}.

It was discussed several times in the literature that the
double-photon production of dileptons in the $p p \to p p l^+ l^-$
reaction can be considered as a luminosity monitor for LHC
\cite{luminosity_monitor}.
Recently we have studied the mechanism of dilepton
production in $\gamma p \to l^+ l^- p$ via exchange of 
gluonic ladder \cite{KSS10}.
The same mechanism can be used in proton-proton collisions when
the photon is in the intermediate state and couples to the parent nucleon
through the proton electromagnetic form factor(s).
It is therefore of interest how this mechanism competes with 
the photon-photon mechanism suggested as the luminosity monitor.
We think therefore that the evaluation of the cross section for 
the diffractive exclusive mechanism is very important in this context.
We wish to make first predictions of the cross section 
for the diffractive exclusive mechanism. We will present several
differential distributions in order to understand the
competition of the diffractive mechanism with the QED one.
We will try to identify regions of phase space where the
diffractive mechanism may dominate over the QED mechanism
which can be helpful in its experimental identification.

This paper is organized as follows,
In Sec.II we present a formalism used to calculations of diffractive processes and
results obtained for single and double diffractive Drell-Yan mechanism for
$\sqrt{s}$ = 500, 1960, 14000 GeV energy. 
In Sec.III we recall a formalism which 
was used to calculate the amplitude ${\cal M}(\gamma p \to \gamma^*(q^2) p)$
as well as we present a formalism for the $pp \to p l^+ l^- p$ reaction both via 
photon--photon fusion and via photon--pomeron (pomeron--photon) fusion.
Next, we compare the results for both contributions.
The last section summarizes our paper.

\section{Inclusive diffractive production of dileptons}

\subsection{Formalism}

\begin{figure}[!h]    %
\includegraphics[width=0.35\textwidth]{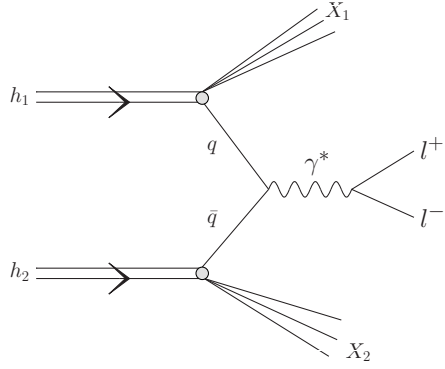}
   \caption{\label{fig:dy_in}
   \small The ordinary leading-order Drell-Yan mechanism of the lepton
   pair production.  
}
\end{figure}

\begin{figure}[!h]    %
\includegraphics[width=0.35\textwidth]{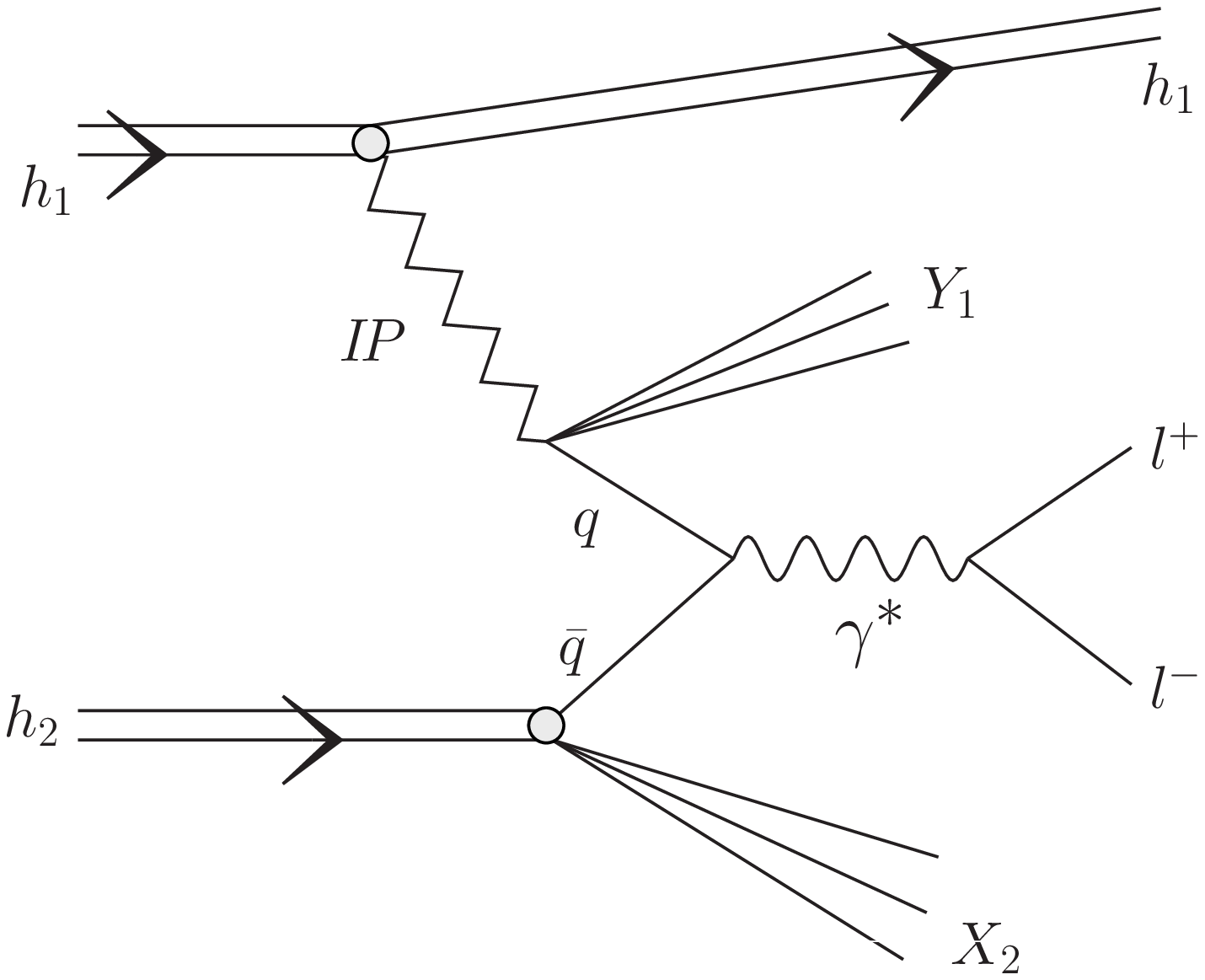}
\hspace{0.5cm}
\includegraphics[width=0.35\textwidth]{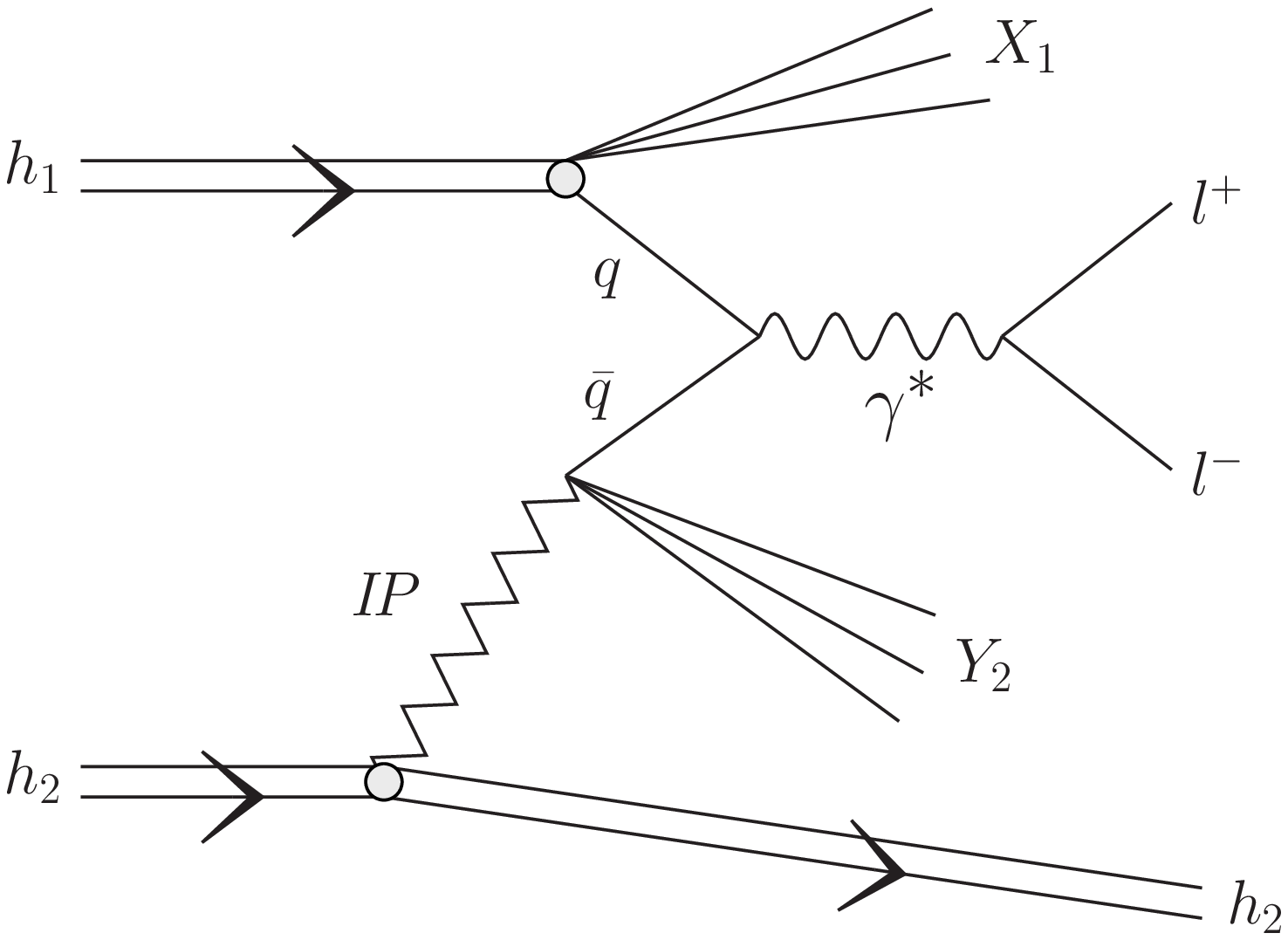}
   \caption{\label{fig:dy_sd}
   \small The mechanism of single-diffractive production of dileptons.  
}
\end{figure}

\begin{figure}[!h]    %
\includegraphics[width=0.35\textwidth]{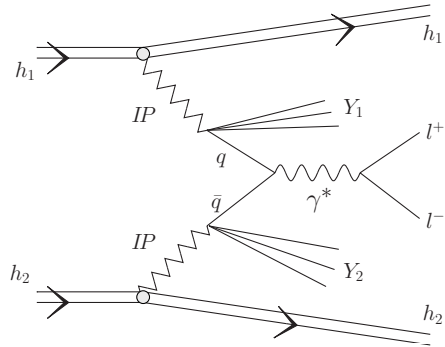}
   \caption{\label{fig:dy_dd}
   \small The mechanism of central-diffractive production of dileptons.  
}
\end{figure}

The mechanisms of the ordinary as well as diffractive production 
of dileptons are shown in Figs.\ref{fig:dy_in},\ref{fig:dy_sd},\ref{fig:dy_dd}.
 

In the following we apply the Ingelman and Schlein approach 
\cite{diffractive_dijets} \footnote{In the literature
also dipole model was used to estimate diffractive processes \cite{Kopeliovich}. }.
In this approach one assumes that the Pomeron has a
well defined partonic structure, and that the hard process
takes place in a Pomeron--proton or proton--Pomeron (single diffraction) 
or Pomeron--Pomeron (central diffraction) processes.
We calculate triple differential distributions
\begin{eqnarray}
{d \sigma_{DY} \over dy_{1} dy_{2} dp_{t}^2} =  K {\Big| M \Big|^2 \over 16 \pi^2 \hat{s}^2} 
\,\Big [\, \Big( x_1 q_f(x_1,\mu^2) 
\, x_2 \bar q_f(x_2,\mu^2) \Big) \, 
+ \Big( x_1 \bar q_f(x_1,\mu^2)
\, x_2  q_f(x_2,\mu^2) \Big) \, \Big ] ,
\nonumber \\ 
\label{DY}
\end{eqnarray}
\begin{eqnarray}
{d \sigma_{SD} \over dy_{1} dy_{2} dp_{t}^2} =  K {\Big| M \Big|^2 \over 16 \pi^2 \hat{s}^2} 
\,\Big [\, \Big( x_1 q_f^D(x_1,\mu^2) 
\, x_2 \bar q_f(x_2,\mu^2) \Big) \, 
+ \Big( x_1 \bar q_f^D(x_1,\mu^2)
\, x_2  q_f(x_2,\mu^2) \Big) \, \Big ] ,
\nonumber \\ 
\label{SD}
\end{eqnarray}
\begin{eqnarray}
{d \sigma_{CD} \over dy_{1} dy_{2} dp_{t}^2} =  K {\Big| M \Big|^2 \over 16 \pi^2 \hat{s}^2} 
\,\Big [\, \Big( x_1 q_f^D(x_1,\mu^2) 
\, x_2 \bar q_f^D(x_2,\mu^2) \Big) \, 
+ \Big( x_1 \bar q_f^D(x_1,\mu^2)
\, x_2  q_f^D(x_2,\mu^2) \Big) \,\Big ] 
\nonumber \\ 
\label{DD}
\end{eqnarray}
for ordinary, single-diffractive and central-diffractive production, respectively.
The matrix element squared for the $q \bar q \to l^{+} l^{-}$
process reads
\begin{eqnarray}
\Big| M  (q \bar q \to l^{+} l^{-})\Big|^2 = 32 \pi^2 \alpha_{em}^2 
{(m_{l}^2 - \hat{t})^2 + (m_{l}^2 - \hat{u})^2 + 2 m_{l}^2 \hat{s} \over \hat{s}^2}.
\nonumber
\end{eqnarray}
In this approach longitudinal momentum fractions are calculated as
\begin{eqnarray}
x_1 = {m_{t} \over \sqrt{s}}  \Big( e^{y_{1}} + e^{y_{2}} \Big) , \\
x_2 = {m_{t} \over \sqrt{s}}  \Big( e^{-y_{1}} + e^{-y_{2}} \Big)
\nonumber
\end{eqnarray}
with $m_{t} = \sqrt{ (p_{t}^2 + m_{l}^2)} \approx p_{t} $.
The distribution in the dilepton invariant mass can be 
obtained by binning differential cross section in $M_{l^+ l^-}$.

We do not calculate the higher-order Drell-Yan contributions and include them effectively with the help of 
a so-called $K$-factor. We have checked that this procedure is precise enough in the case of ordinary Drell-Yan.
The $K$-factor for the Drell-Yan process can be calculated as \cite{BP87}
\begin{eqnarray*}
K &=& 1 +{ \alpha_{s}  \over  2 \pi}{ 4 \over 3} \Big (1+ { 4 \over 3} \pi^2 \Big).
\end{eqnarray*}
Here the running coupling constant $\alpha_{s} = \alpha_{s}(\mu^2)$ is
evaluated at $\mu^2 = M_{l^+l^-}^2$.

The 'diffractive' quark distribution of
flavour $f$ can be obtained by a convolution of the flux of Pomerons
$f_\Pom(x_\Pom)$ and the parton distribution in the Pomeron 
$q_{f/\Pom}(\beta, \mu^2)$:
\begin{eqnarray}
q_f^D(x,\mu^2) = \int d x_\Pom d\beta \, \delta(x-x_\Pom \beta) 
q_{f/\Pom} (\beta,\mu^2) \, f_\Pom(x_\Pom) \, 
= \int_x^1 {d x_\Pom \over x_\Pom} \, f_\Pom(x_\Pom)  
q_{f/\Pom}({x \over x_\Pom}, \mu^2) \, . \nonumber \\
\end{eqnarray}
The flux of Pomerons $f_\Pom(x_\Pom)$ enters in the form integrated over 
four--momentum transfer 
\begin{eqnarray}
f_\Pom(x_\Pom) = \int_{t_{min}}^{t_{max}} dt \, f(x_\Pom,t) \, ,
\label{flux_of_Pom}
\end{eqnarray}
with $t_{min}, t_{max}$ being kinematic boundaries.

Both pomeron flux factors $f_{\Pom}(x_{\Pom},t)$ as well 
as quark/antiquark distributions in the pomeron were taken from 
the H1 collaboration analysis of diffractive structure function
and diffractive dijets at HERA\cite{H1}. 
The factorization scale for diffractive parton distributions is taken as
$\mu^2 = M_{ll}^2$.

\subsection{Absorption corrections}

Up to now we have assumed Regge factorization which is known
to be violated in hadron-hadron collisions.
It is known that soft interactions lead to an extra production
of particles which fill in the rapidity gaps related to pomeron
exchange.

Different models of absorption corrections 
(one-, two- or three-channel approaches) 
for diffractive processes were presented in the literature.
The absorption effects for the diffractive processes were calculated e.g.
in \cite{KMR2000,Maor2009,CSS09}.
The different models give slightly different predictions.
Usually an average value of the gap survival probability
$<|S|^2>$ is calculated first and then the cross sections for different
processes is multiplied by this value.
We shall follow this somewhat simplified approach also here.
Numerical values of the gap survival probability can be found 
in \cite{KMR2000,Maor2009,CSS09}.
The survival probability depends on the collision energy.
It is sometimes parametrized as:
\begin{equation}
<|S|^2>(\sqrt{s}) = \frac{a}{b+\ln(\sqrt{s})} \; .
\end{equation}
The numerical values of the parameters can be found in original
publications.
As discussed in \cite{CSS09,KMR2000} the absorptive corrections for single and
central difractive DY are somewhat different.

\subsection{Results}
\begin{figure}[!ht]    %
\begin{center}
\includegraphics[width=0.3\textwidth]{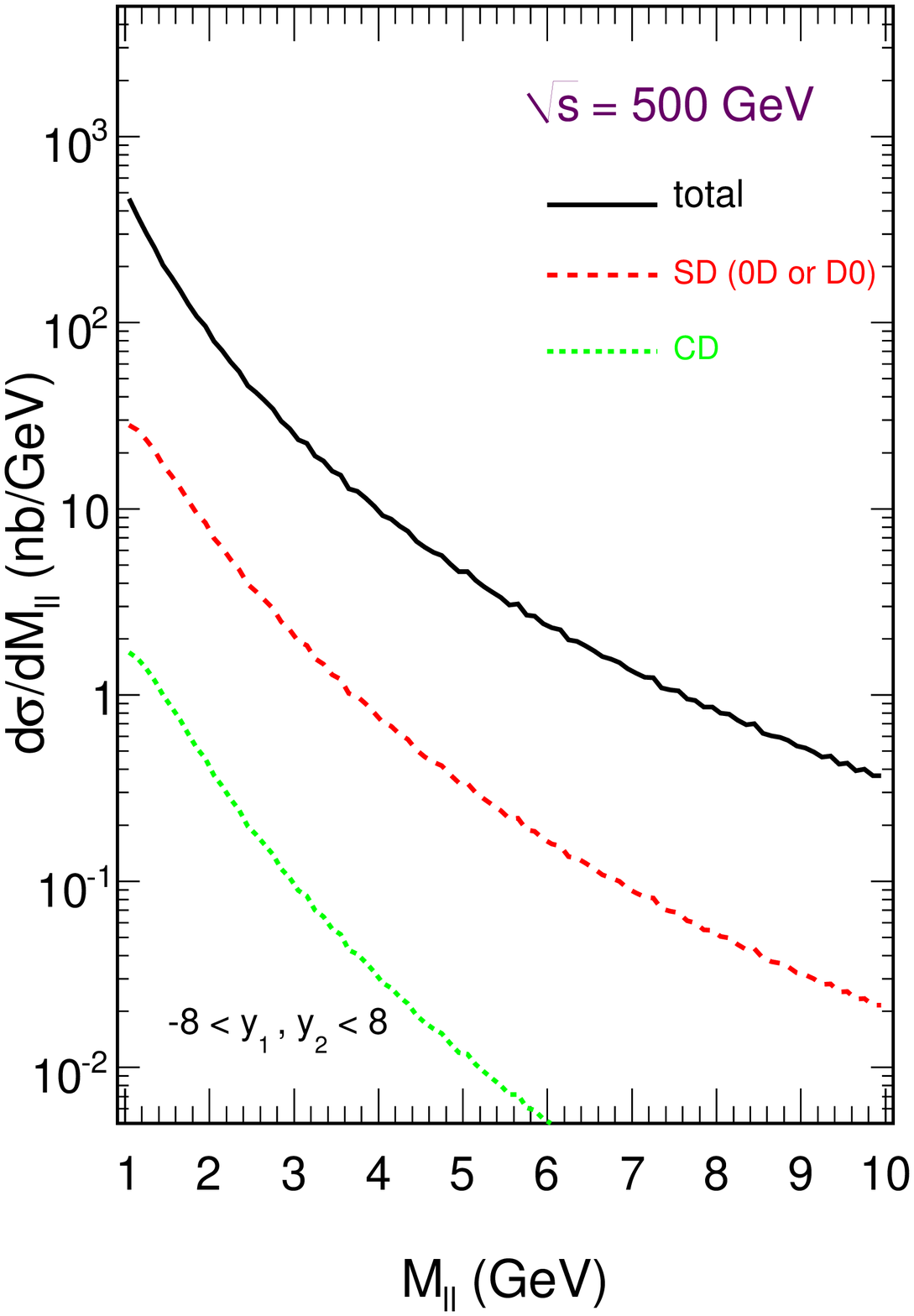}
\includegraphics[width=0.3\textwidth]{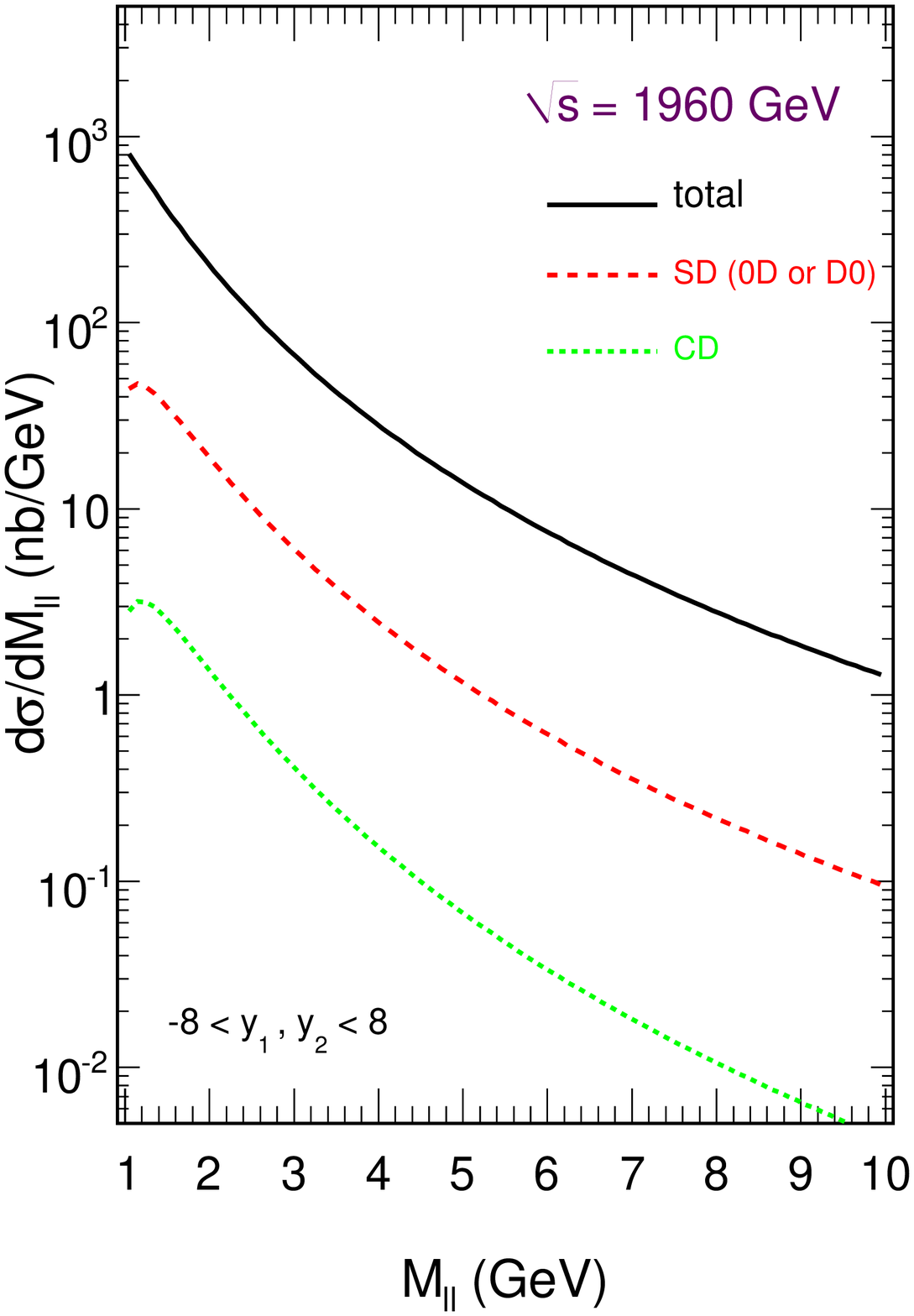}
\includegraphics[width=0.3\textwidth]{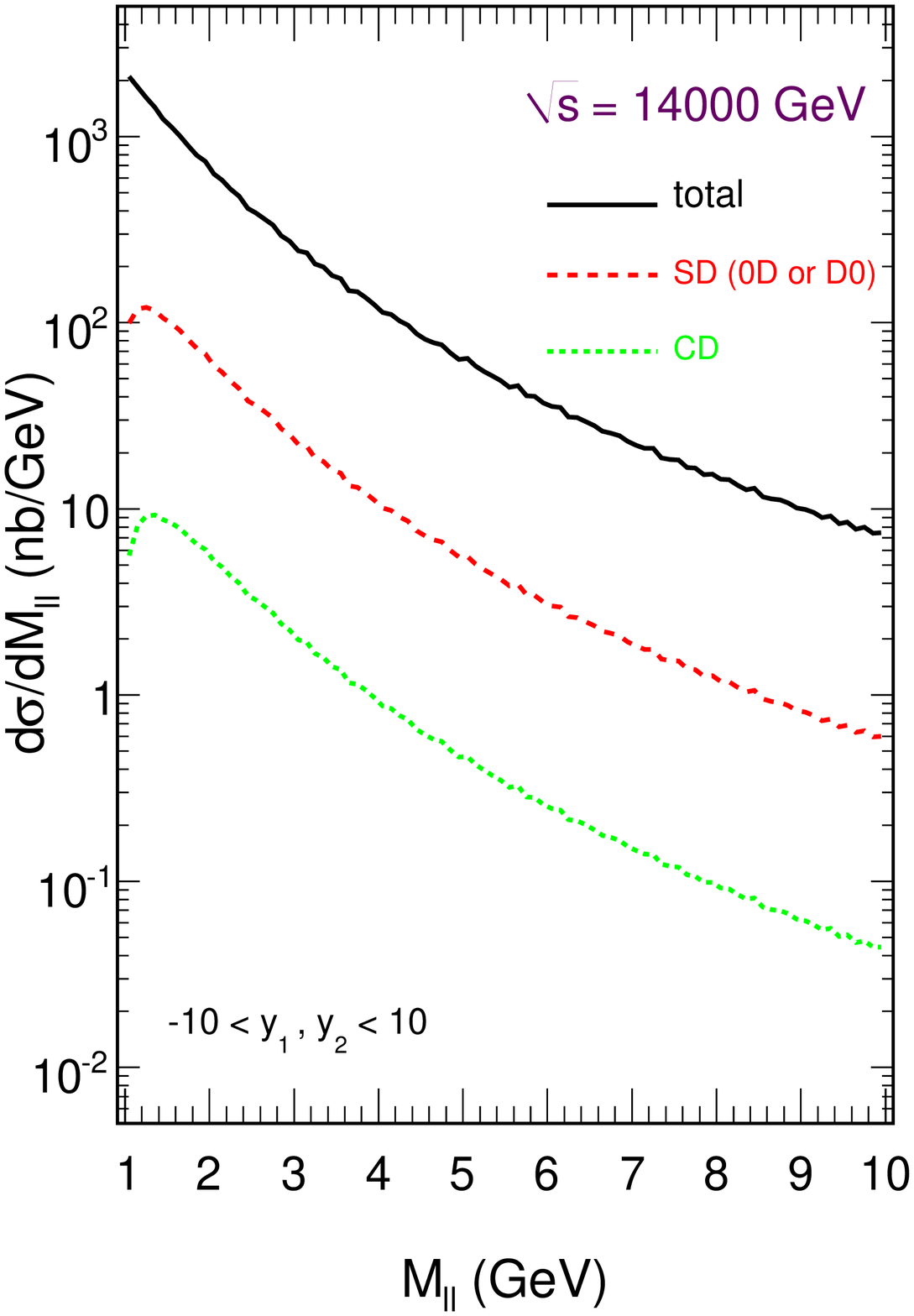}
\end{center}
   \caption{\label{fig:m_ll_diff}
   \small
Invariant mass ($M_{ll}$) distributions for the ordinary Drell-Yan (black line),
single diffractive DY (dashed, red online) and central diffractive DY (dotted green online).
The results are shown for energies $\sqrt{s}$= 500, 1960, 14000 GeV and the full lepton rapidity
interval.
}
\end{figure}

%
In this section we shall present several differential distributions
for the inclusive diffractive production of the dilepton pairs.
Let us start from the presentation of one-dimensional
distributions.\\
In the presentation below we shall show results for dimuon production.
The cross sections for dielectrons are larger within the line thickness.
In Fig.\ref{fig:m_ll_diff} we show invariant mass 
distributions of dilepton pair.
We compare contributions of diffractive and ordinary Drell-Yan processes. 
The single diffractive distributions are smaller than that for the ordinary
Drell-Yan by a factor 10. The calculation done assumes Regge factorization.
Absorption corrections, i.e. Regge factorization violation, can be taken
into account by a multiplicative factor being a probability of a rapidity gap survival
(see e.g.\cite{CSS09}). Such a factor is approximately $S_G$ = 0.2 for the RHIC energy $\sqrt{s}
=$ 500 GeV, $S_G$ = 0.1 for the Tevatron energy $\sqrt{s}=$ 1960 GeV and $S_G$ = 0.05
for the LHC energy $\sqrt{s}=$ 14 TeV. 
The diffractive distributions shown should be multiplied in addition by these factors.
In order to avoid model dependence we shall include them only
when comparing diffrent contributions (the reader can use his/her own numbers).

\begin{figure}[!ht]    %
\begin{center}
\includegraphics[width=0.3\textwidth]{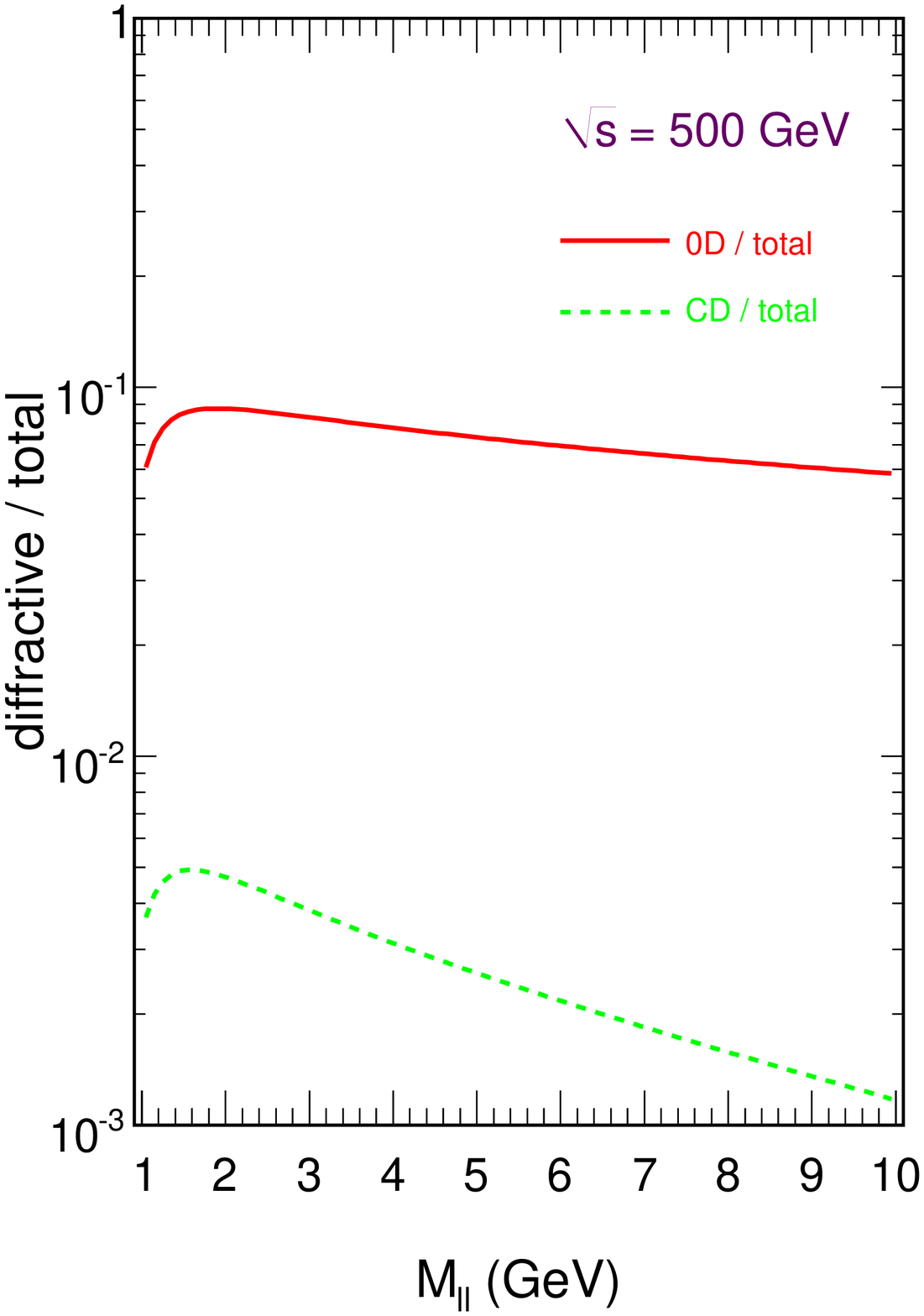}
\includegraphics[width=0.3\textwidth]{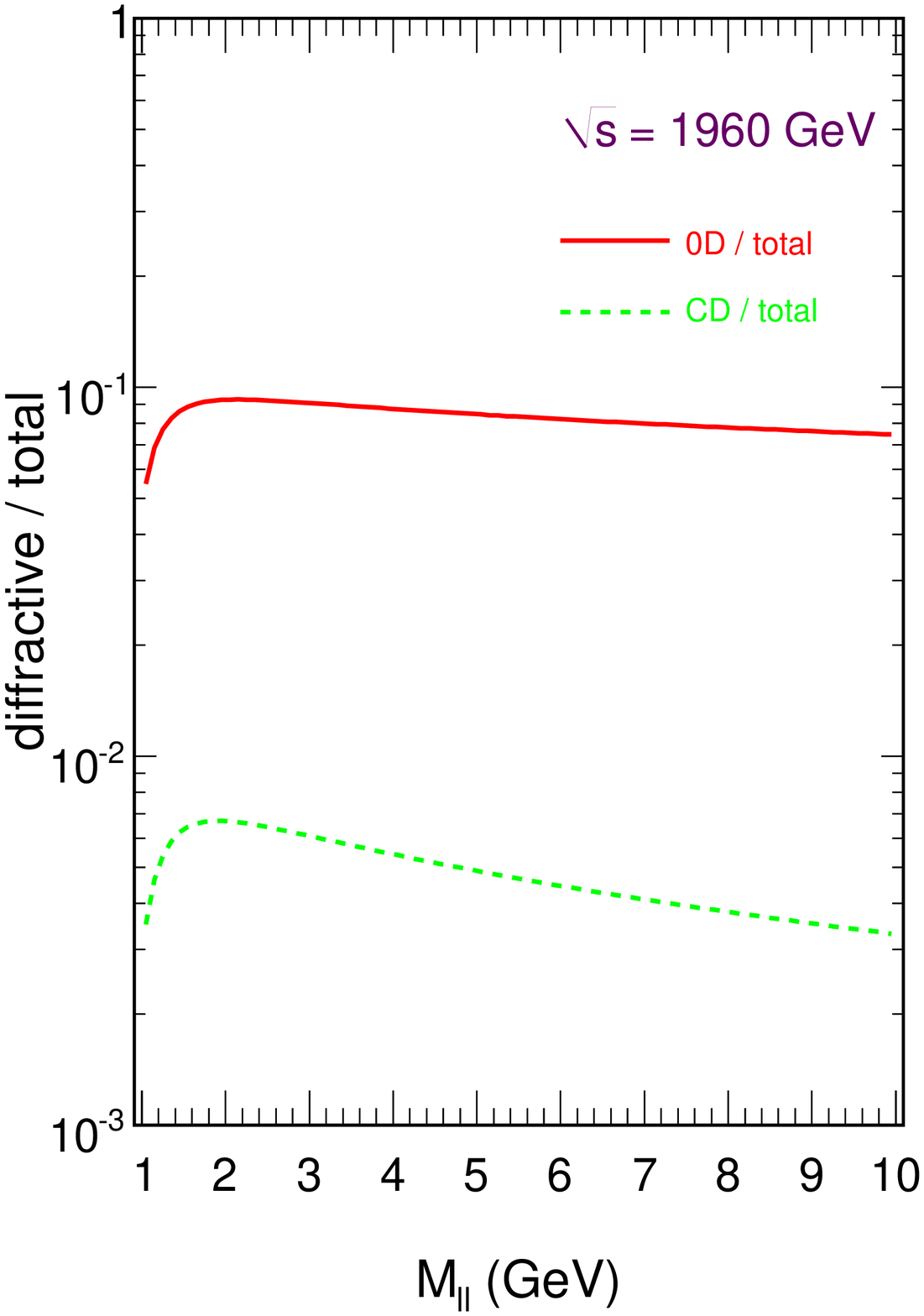}
\includegraphics[width=0.3\textwidth]{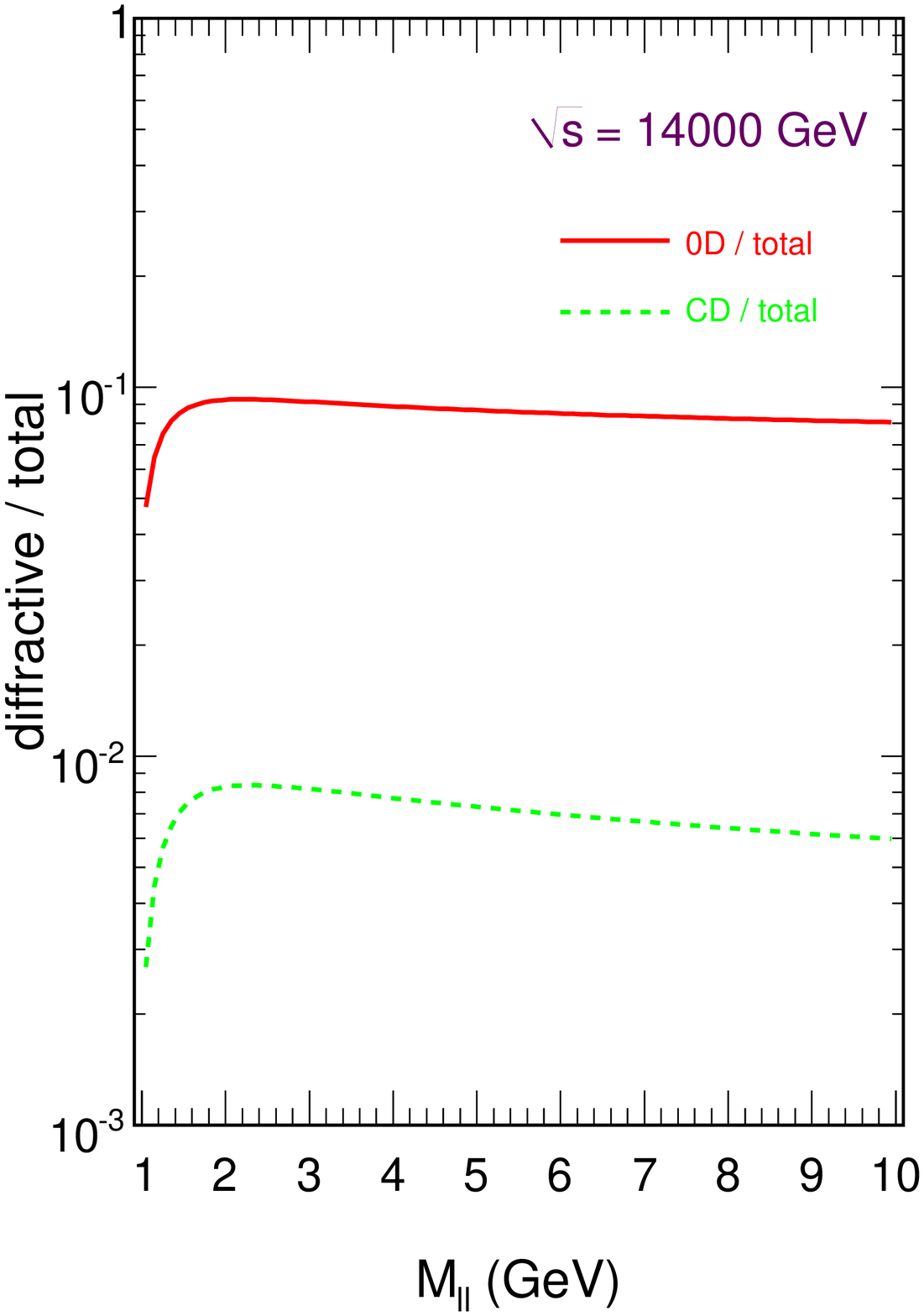}
\end{center}
   \caption{\label{fig:ratio}
   \small
Ratio of the single-diffractive DY (solid line) and central-diffractive DY (dashed line)
cross section to the ordinary Drell-Yan cross section as a function of the dilepton
invariant mass. The gap survival factors are not included in this plot and have to
be taken into account in addition.
}
\end{figure}

In Fig.\ref{fig:ratio} we show the ratios of the cross sections of diffractive (single and central) 
to the ordinary Drell-Yan for different energies $\sqrt{s}$= 500, 1960, 14000 GeV.
We did not include here the gap survival factors which are model dependent.
The ratios for single-diffractive DY is almost independent of dilepton mass whereas
that for central-diffractive DY slightly decreases with the dilepton mass. The result found here differs
from that found in the dipole approach in Ref.\cite{Kopeliovich}. 
In addition the ratio for the single-diffractive component is almost
energy independent (please note that here we have not included the gap survival
factor which decreases logarithmically with energy).
This evident difference between our approach and the dipole approach requires a deeper understanding 
in the future.

\begin{figure}[!ht]    %
\begin{center}
\includegraphics[width=0.3\textwidth]{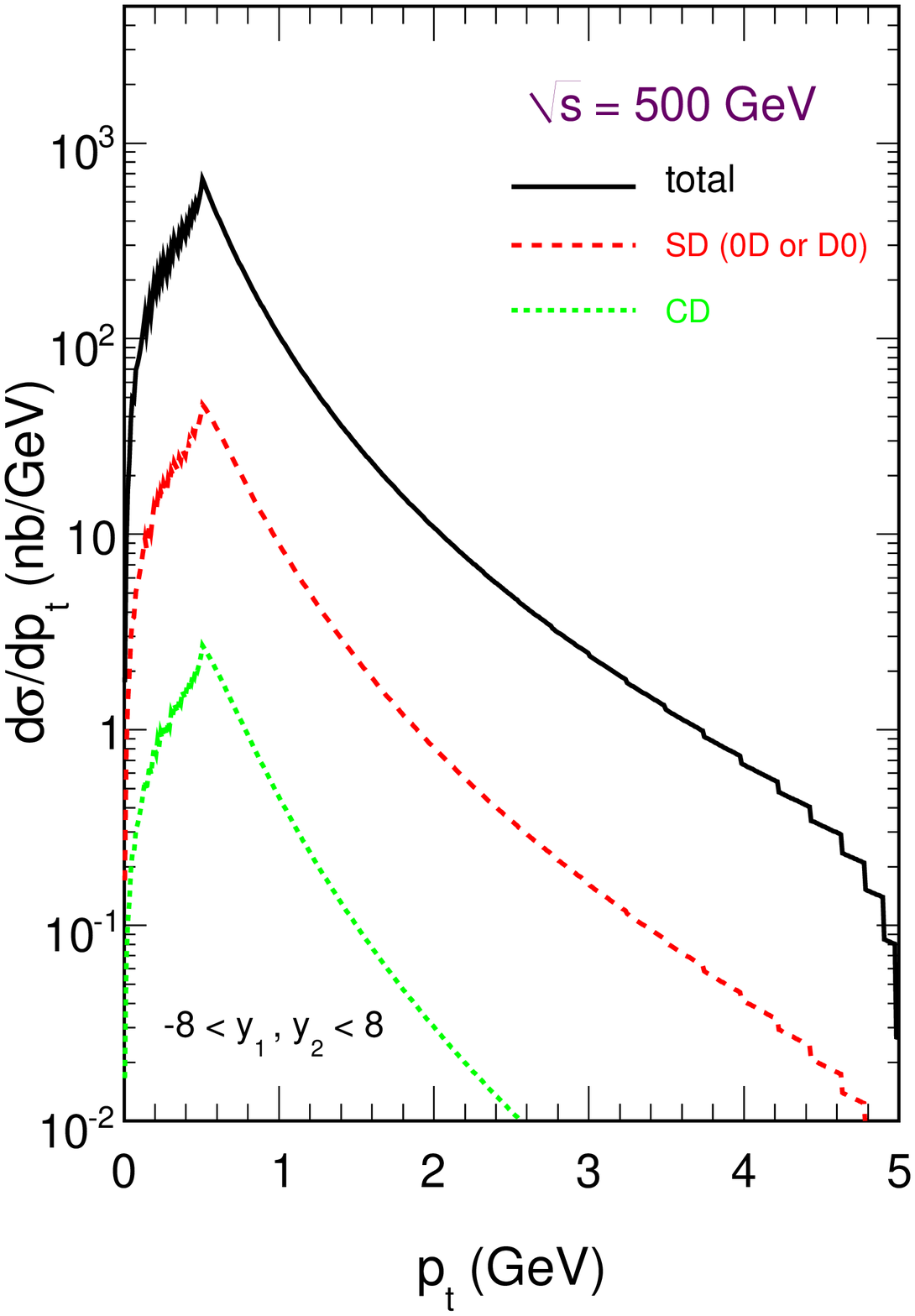}
\includegraphics[width=0.3\textwidth]{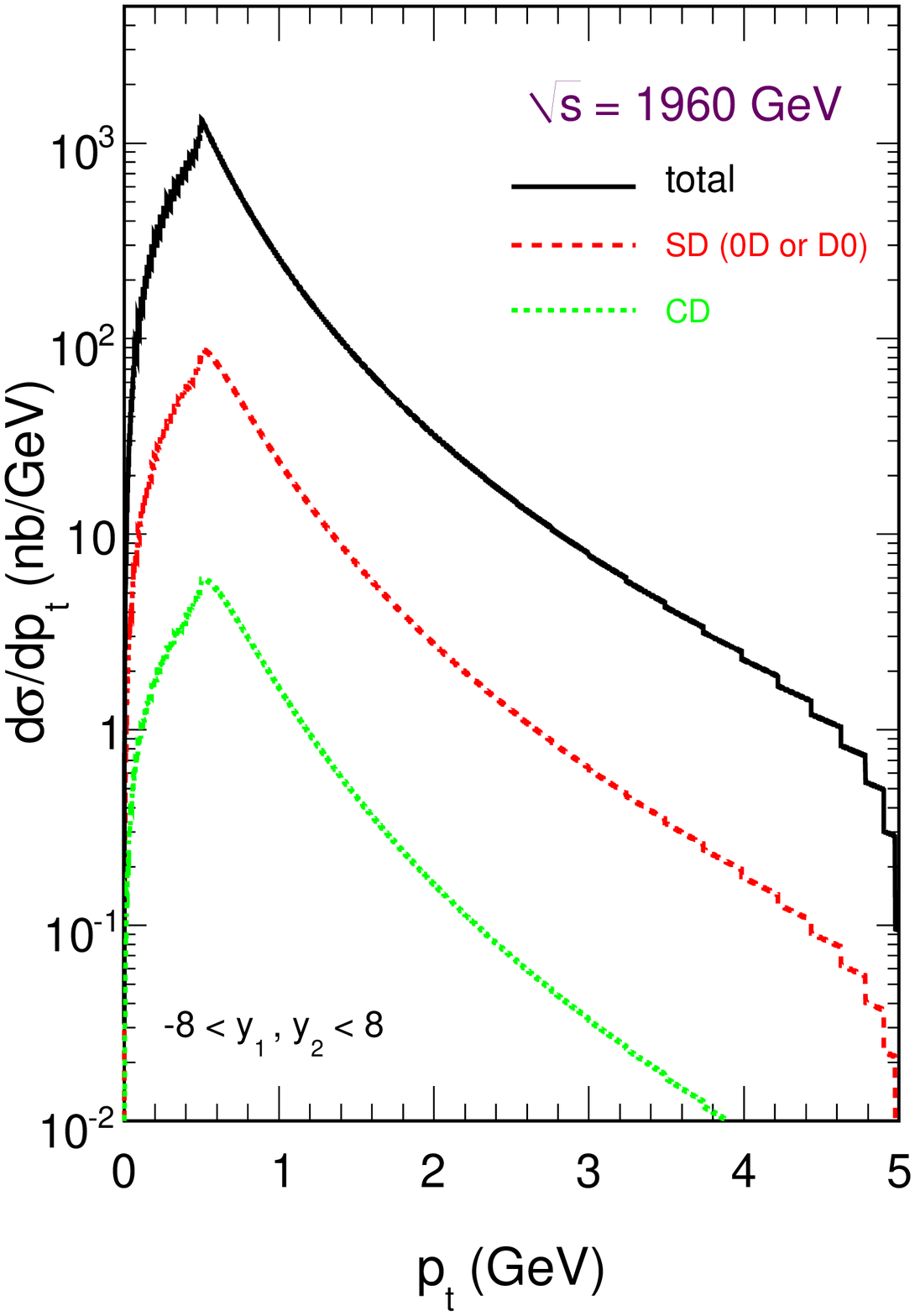}
\includegraphics[width=0.3\textwidth]{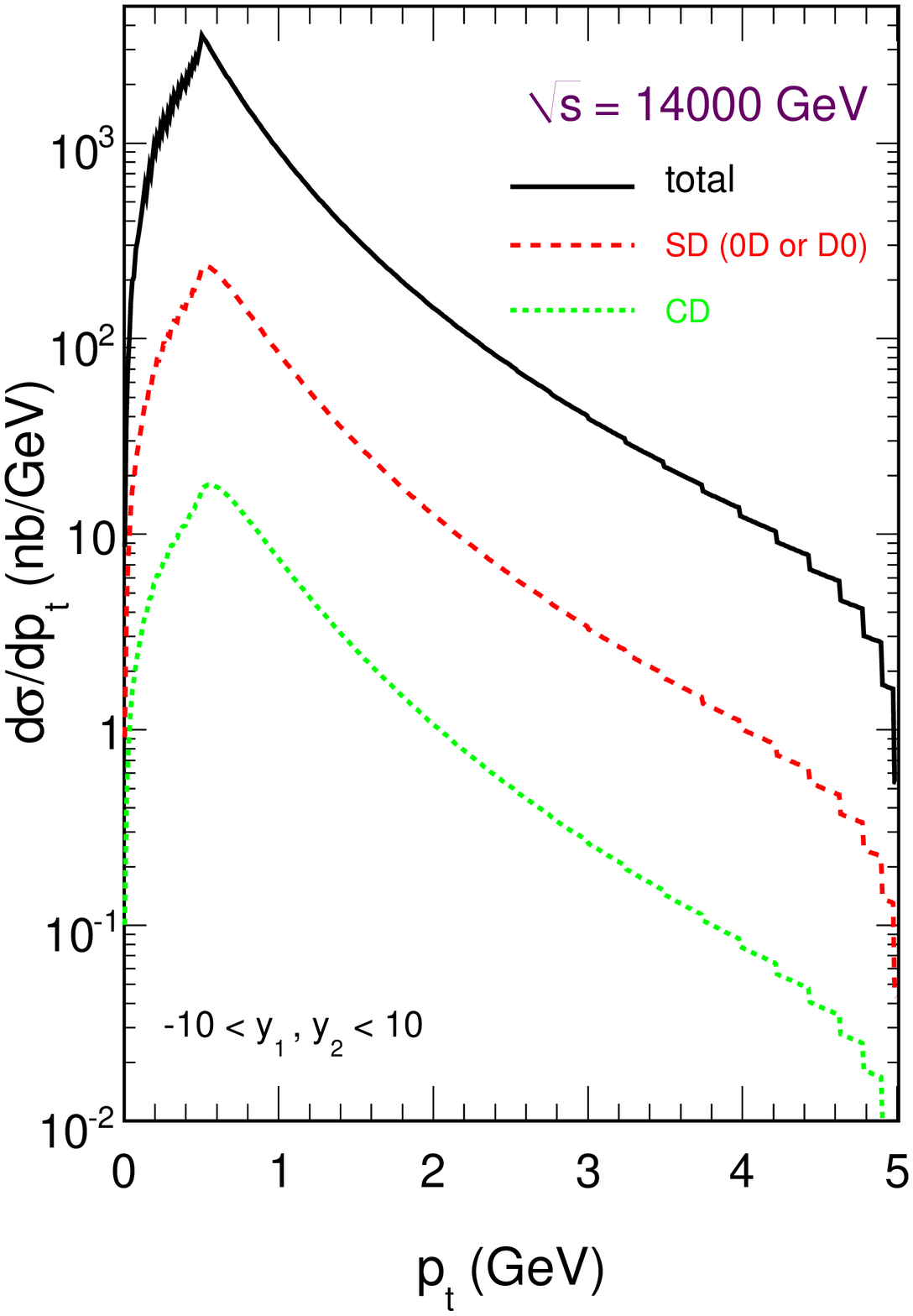}
\end{center}
   \caption{\label{fig:pt_diff}
   \small
Lepton transverse momentum distributions for the ordinary Drell-Yan (black line),
single diffractive DY (dashed, red online) and central diffractive DY (dotted, green online).
The results are shown for energies $\sqrt{s}$= 500, 1960, 14000 GeV and the full lepton rapidity
interval. Absorption effects are not included here.
}
\end{figure}

In Fig.\ref{fig:pt_diff} we show distribution in transverse momentum
of individual leptons for diffractive contributions at $\sqrt{s}$ = 500,
1960, 14000 GeV.
For comparison we show also prediction for the ordinary Drell-Yan
contribution.
A somewhat strange shape for $p_{t} \in $ (0-1) GeV is a consequence of the
cut imposed on the dilepton invariant mass $M_{ll} >$ 1 GeV, necessary to ensure 
validity of the perturbative calculation.

\begin{figure}[!ht]    %
\begin{center}
\includegraphics[width=0.3\textwidth]{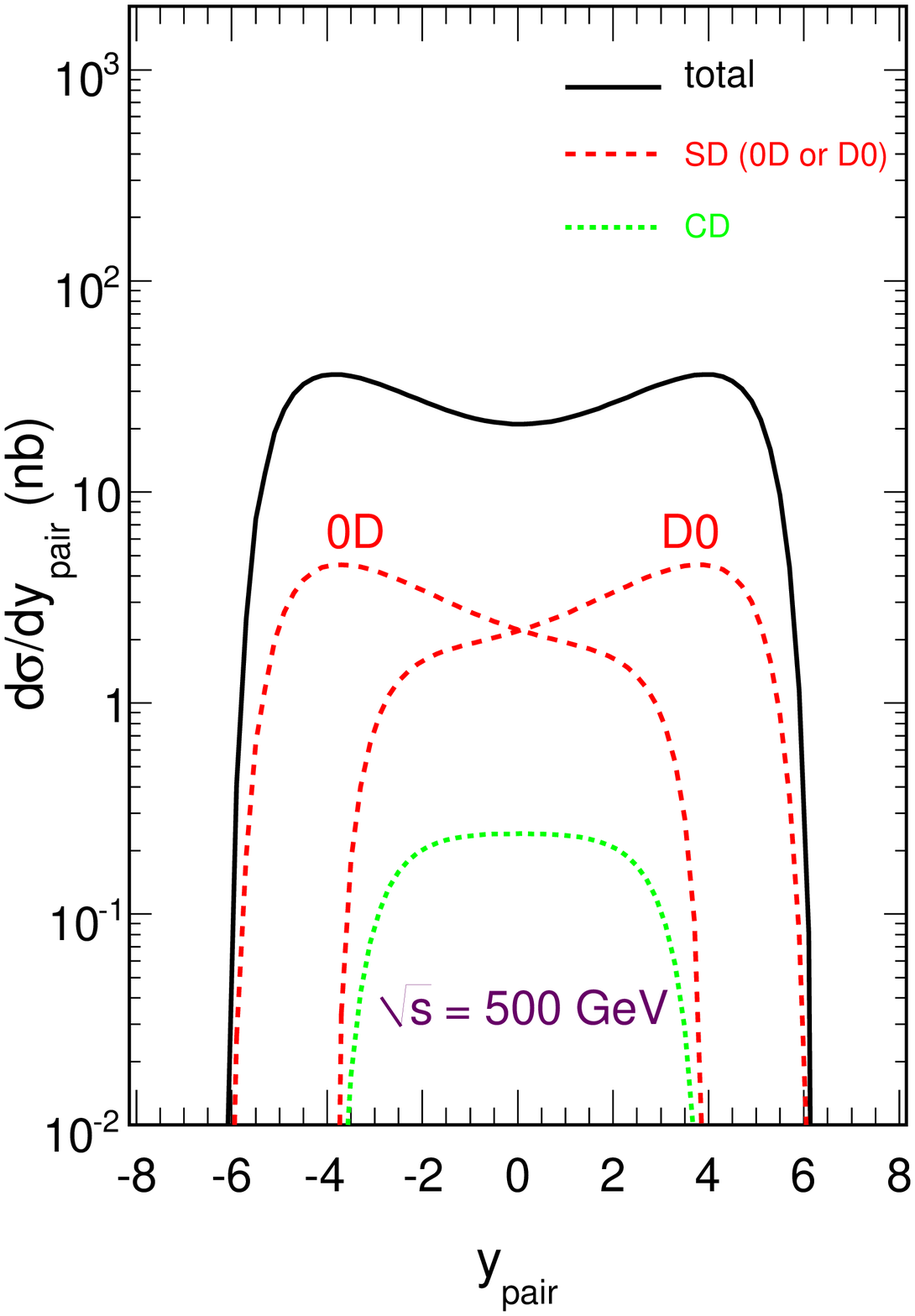}
\includegraphics[width=0.3\textwidth]{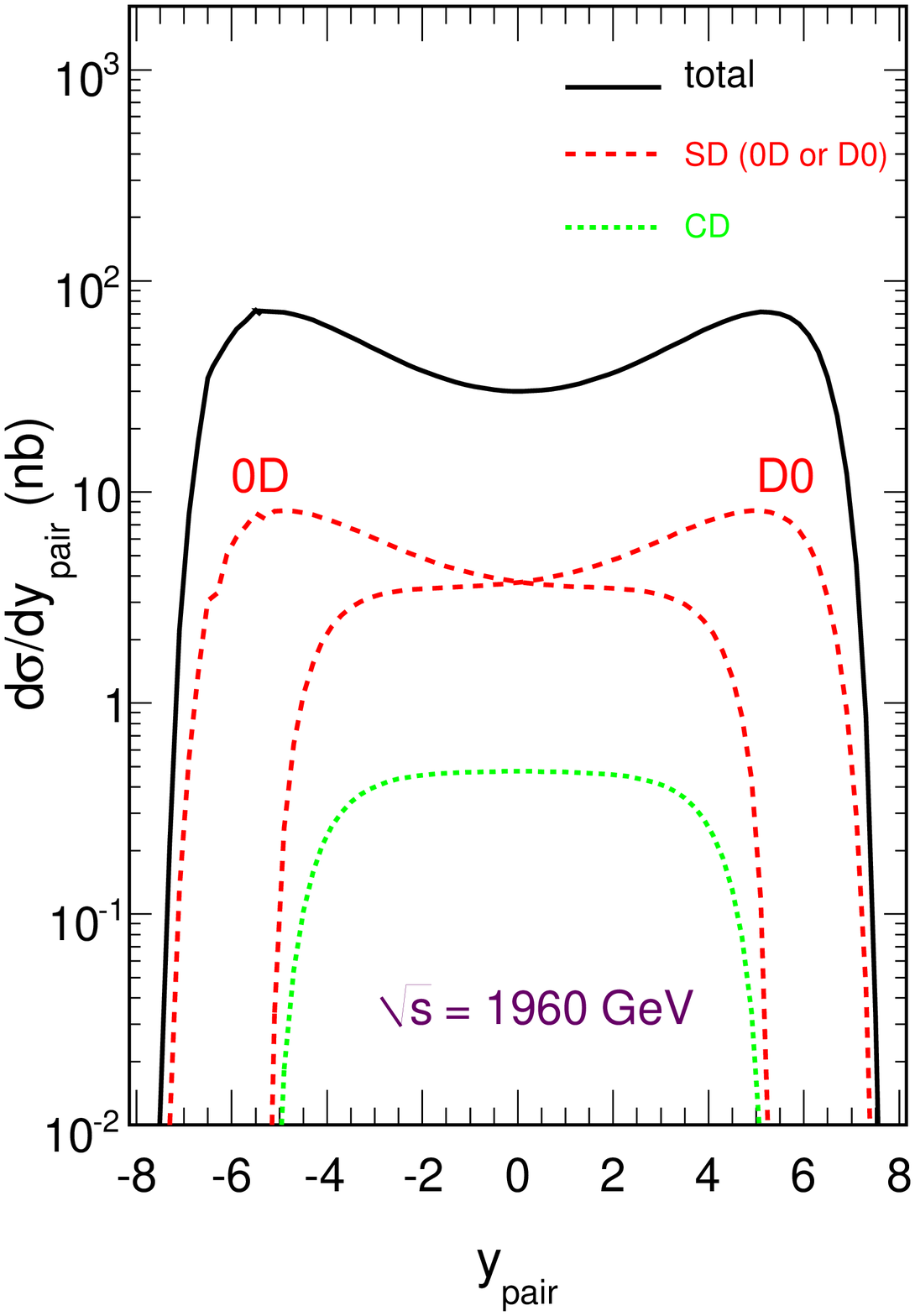}
\includegraphics[width=0.3\textwidth]{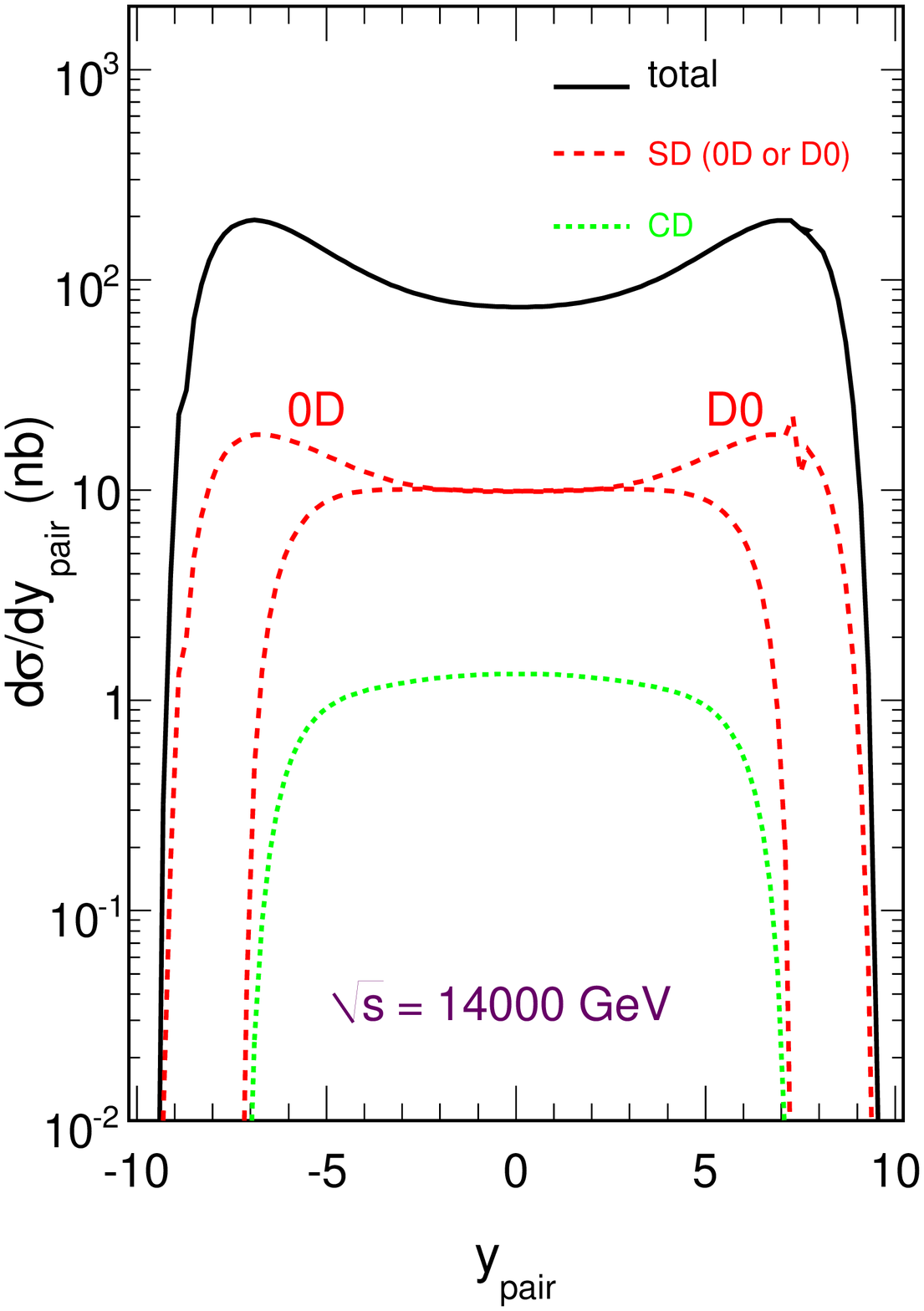}
\end{center}
   \caption{\label{fig:y_pair_diff}
   \small
Distribution in the lepton pair rapidity ($y_{pair}$) for 
$\sqrt{s}$= 500, 1960, 14000 GeV energy.
Here $M_{ll} > $ 1 GeV. Absorption effects are not included here.
}
\end{figure}

The rapidity distribution of the dilepton pair is shown in
Fig.\ref{fig:y_pair_diff}. The distributions for the individual single diffractive
mechanisms have maxima at large rapidities. When adding both single
diffractive contributions we obtain distribution which has a shape
similar to that for the ordinary Drell-Yan. This means that the fraction of the
single diffractive contribution is only weakly dependent on the lepton pair 
rapidity. The central diffractive contribution is concentrated
at midrapidities. This is a consequence of limiting integration
over $x_\Pom$ in Eq.(\ref{flux_of_Pom}) to 0.0 $< x_\Pom <$ 0.1 .


\begin{figure}[!ht]    %
\begin{center}
\includegraphics[width=0.4\textwidth]{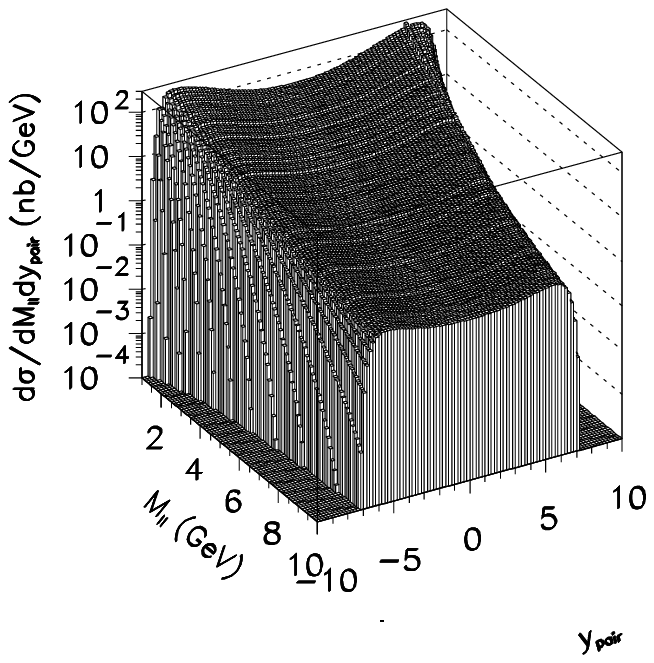}
\includegraphics[width=0.4\textwidth]{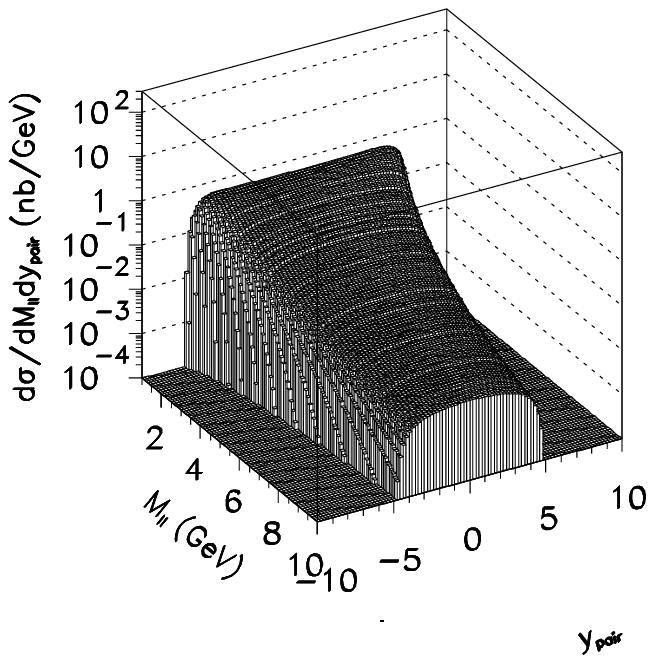}
\includegraphics[width=0.4\textwidth]{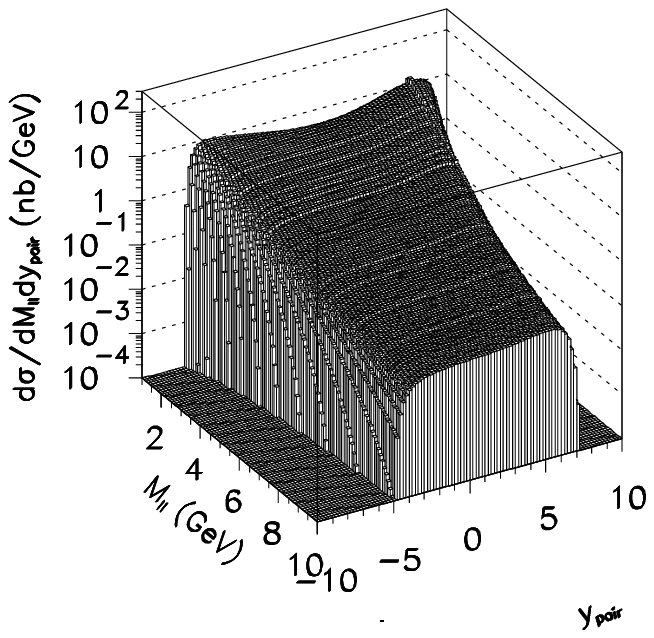}
\includegraphics[width=0.4\textwidth]{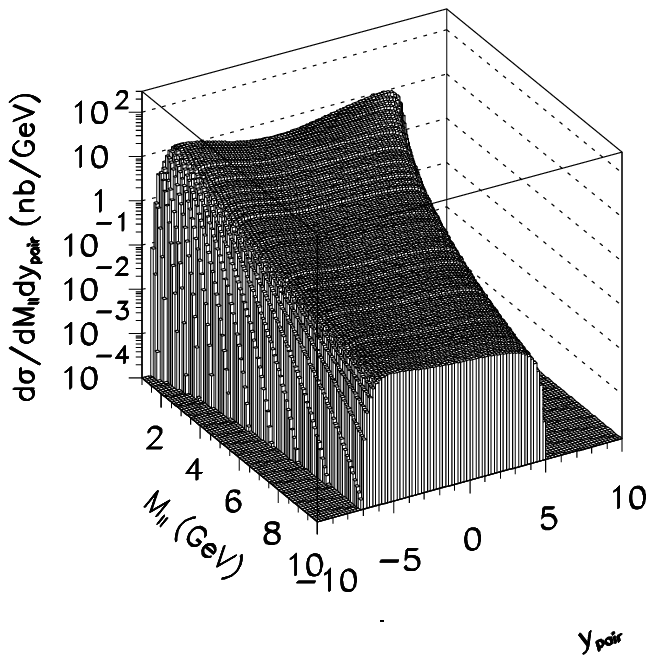}
\end{center}
   \caption{\label{fig:m_ll_y_pair}
   \small
Two-dimensional distributions in dilepton invariant mass $M_{ll}$ and 
lepton pair rapidity $y_{pair} $ for
ordinary Drell-Yan (upper left panel), central diffractive DY (upper right panel)
and single diffractive DY (lower panels) in $pp$ colisions at 
$\sqrt{s}$=14000 GeV. The cross sections for diffractive processes
were not multiplied by the gap survival factors.
}
\end{figure}

Now we wish to discuss some two-dimensional distributions.
In Fig.\ref{fig:m_ll_y_pair} maps in ($M_{ll}$, $y_{pair}$) for the
four diffrent processes are presented.
The shapes for different processes are somewhat different. In particularly
the distribution of the central-diffractive component is
much narrower in the rapidity of the lepton pair.


\begin{figure}[!ht]    %
\begin{center}
\includegraphics[width=0.4\textwidth]{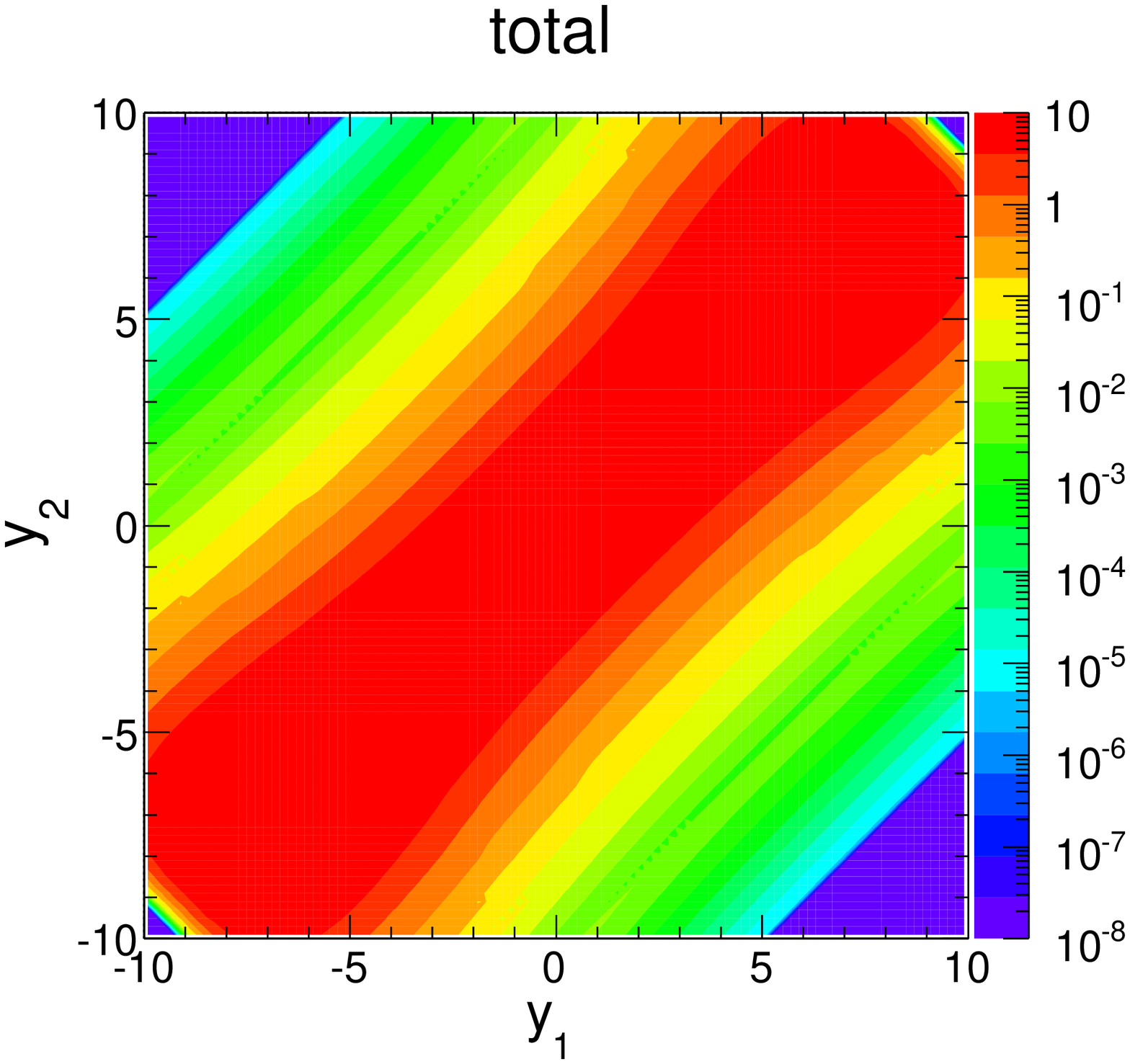}
\includegraphics[width=0.4\textwidth]{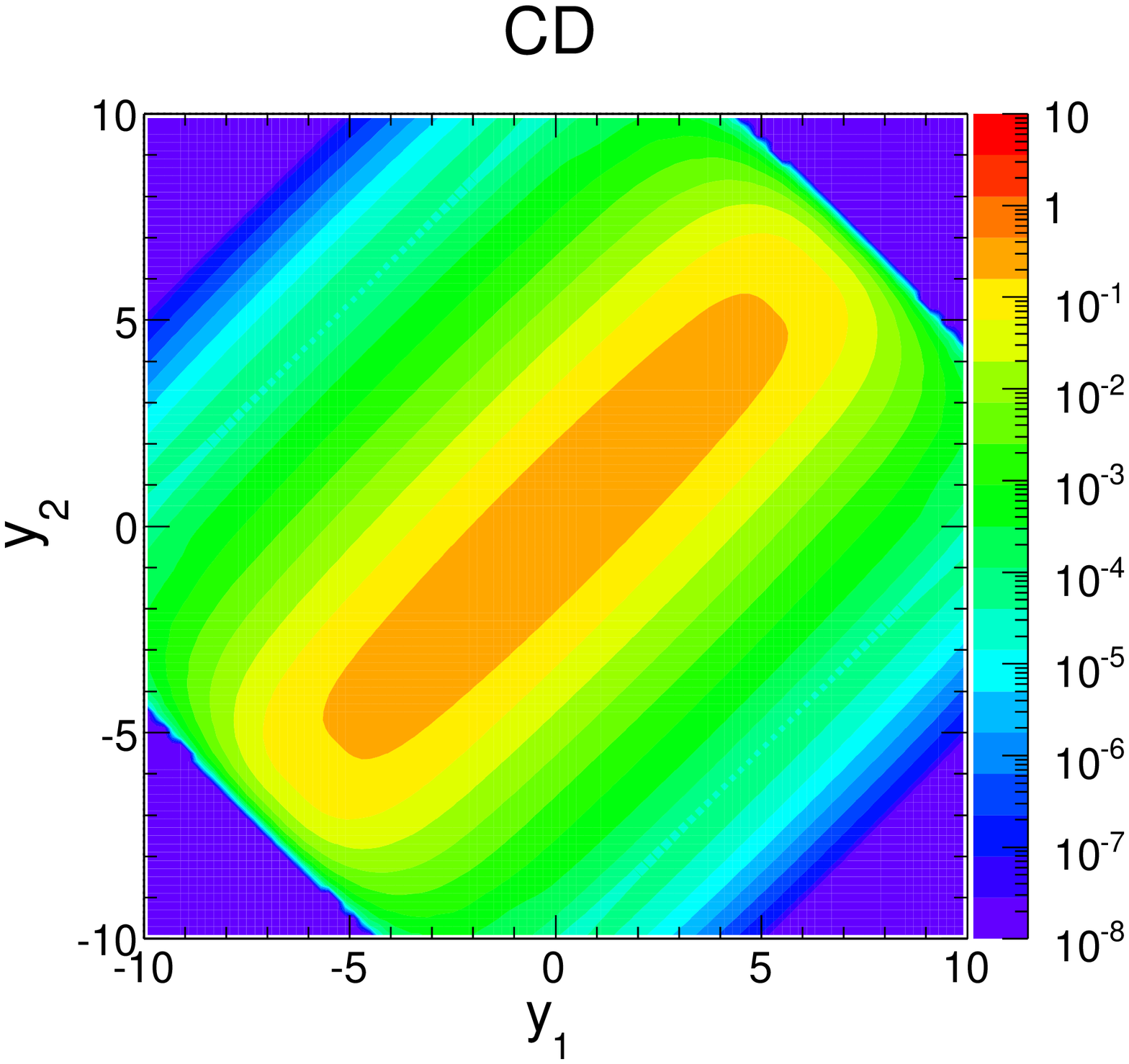}
\includegraphics[width=0.4\textwidth]{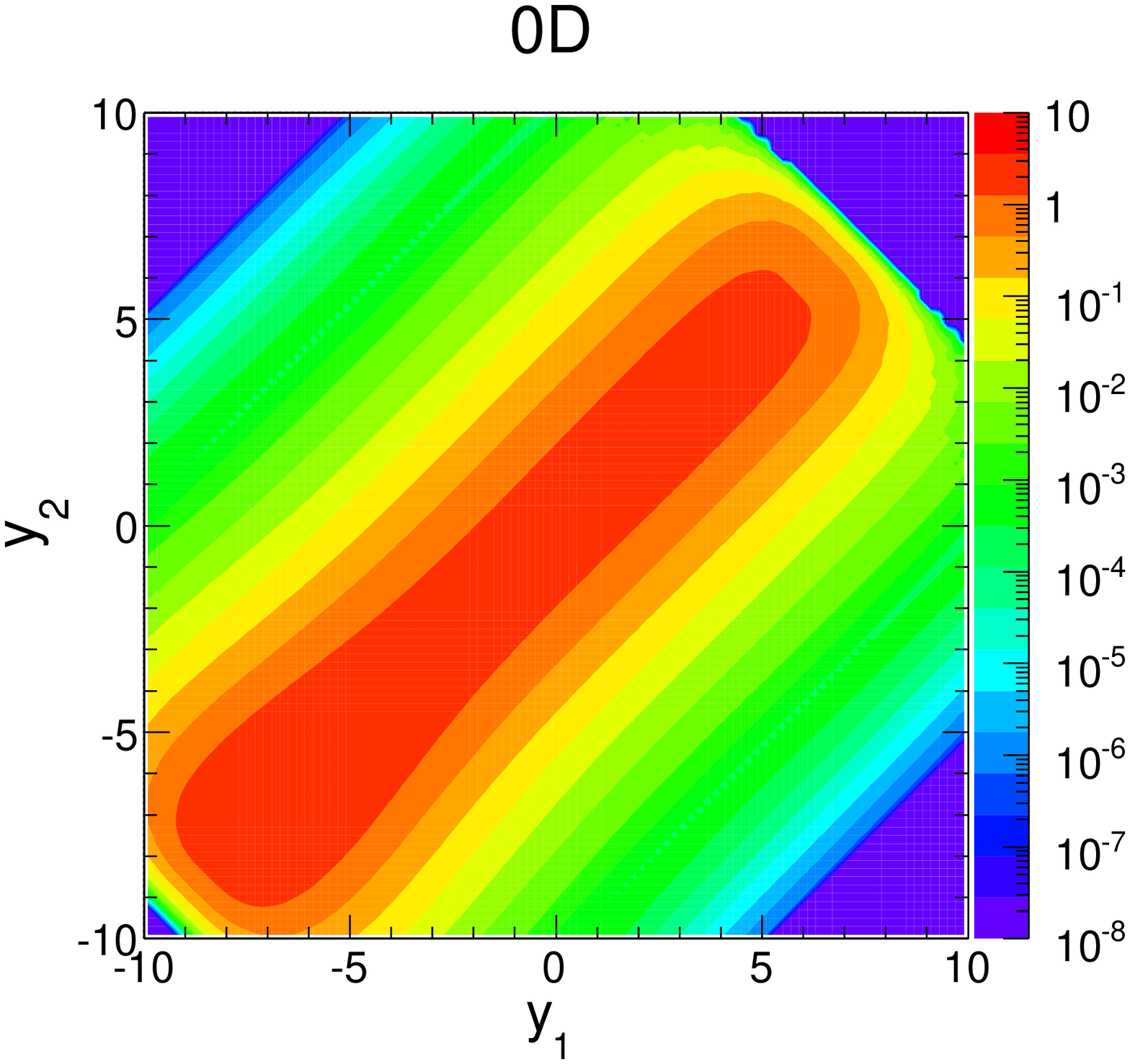}
\includegraphics[width=0.4\textwidth]{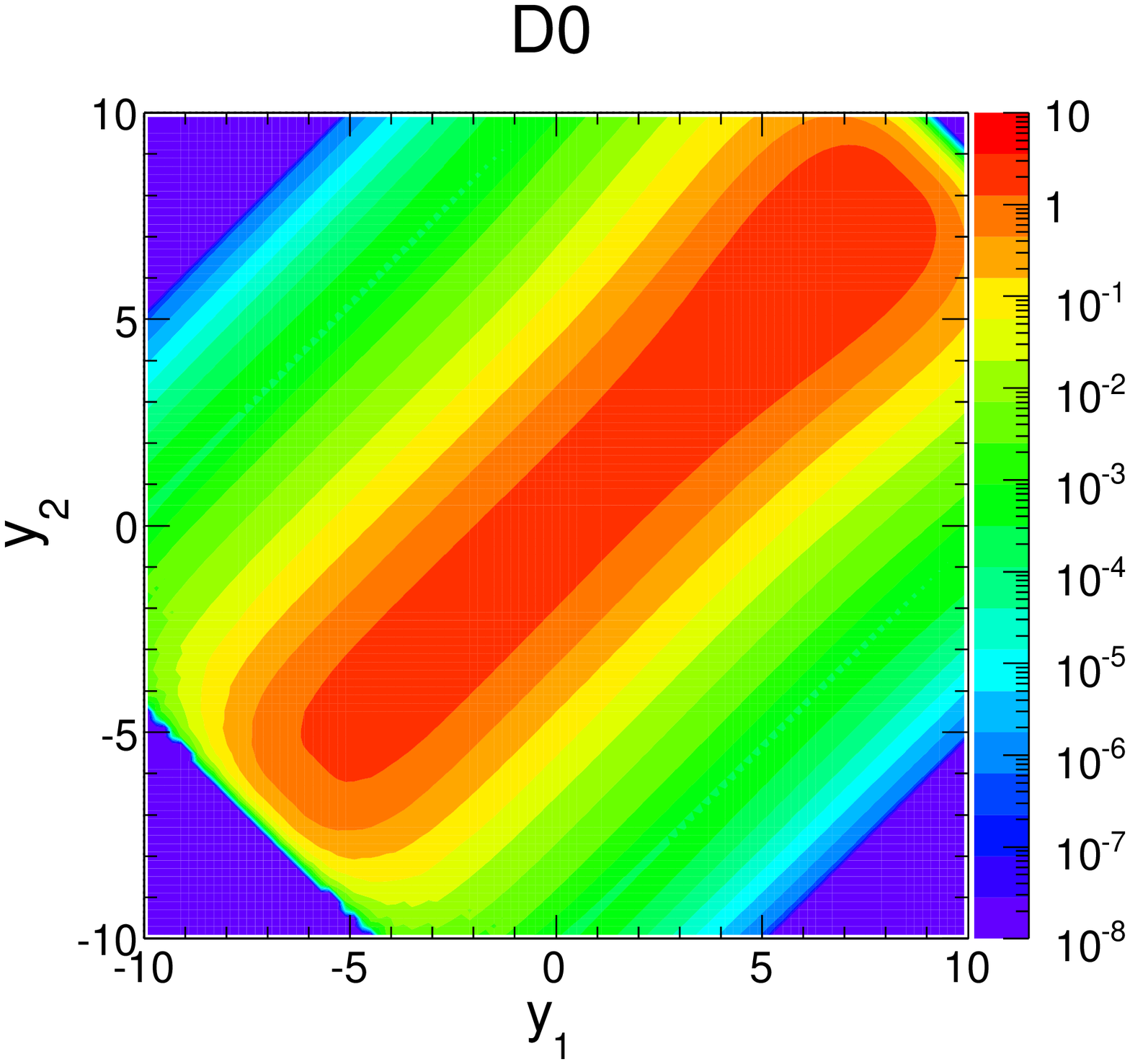}
\end{center}
   \caption{\label{fig:y1_y2}
   \small
Maps in $y_{1}$ and $y_{2} $ for the
ordinary Drell-Yan (upper left panel), central diffractive DY (upper right panel)
and single diffractive DY (lower panels) in $pp$ colisions at 
$\sqrt{s}$=14000 GeV.
The cross sections for the diffractive processes were not multiplied by
the gap survival factors.
}
\end{figure}

In Fig.\ref{fig:y1_y2} we show correlations
in lepton rapidities ($y_1$ is for $e^-$ and $y_2$ is for $e^+$).
The distributions are concentrated along the diagonal $y_1 = y_2$.
The distributions for individual single-diffractive components
are peaked at large $|y_1| \approx |y_2|$. 
The LHC detectors have fairly large coverage in pseudorapidity for leptons
so a measurement of such distributions in the near future is not excluded.


In the present paper we have calculated the cross section
for single diffractive mechansim within a simple intuitive
Ingelman-Schlein model \cite{diffractive_dijets}.
We find that the ratio of the diffractive to the total cross section
for dilepton pair production depends only slightly on kinematical
variables: center-of-mass energy, $M_{ll}$, $y_{pair}$ or 
lepton transverse momenta.
In Ref.\cite{Kopeliovich} such ratio was calculated
in the framework of the dipole approach to the Drell-Yan mechanism
as a function of dilepton invariant mass for different center-of-mass
energies.
In this approach the ratio strongly decreases as a function of 
the center-of-mass energy and increases as a function of dilepton
invariant mass. This is quite opposite to the present predictions,
where the energy dependence of the ratio is rather slow and the ratio rather decreases
as a function of the dilepton invariant mass.

\clearpage

\section{Exclusive production of dileptons}

\subsection{$pp \to p l^+ l^- p$ \, via photon--pomeron subprocesses}

Before we go to hadronic reaction let us start from recalling basic formula
for the amplitude for exclusive photoproduction of lepton pairs in the 
$\gamma p \to l^+ l^- p$ reaction.


\begin{figure}[!ht]    %
\begin{center}
\includegraphics[width=0.4\textwidth]{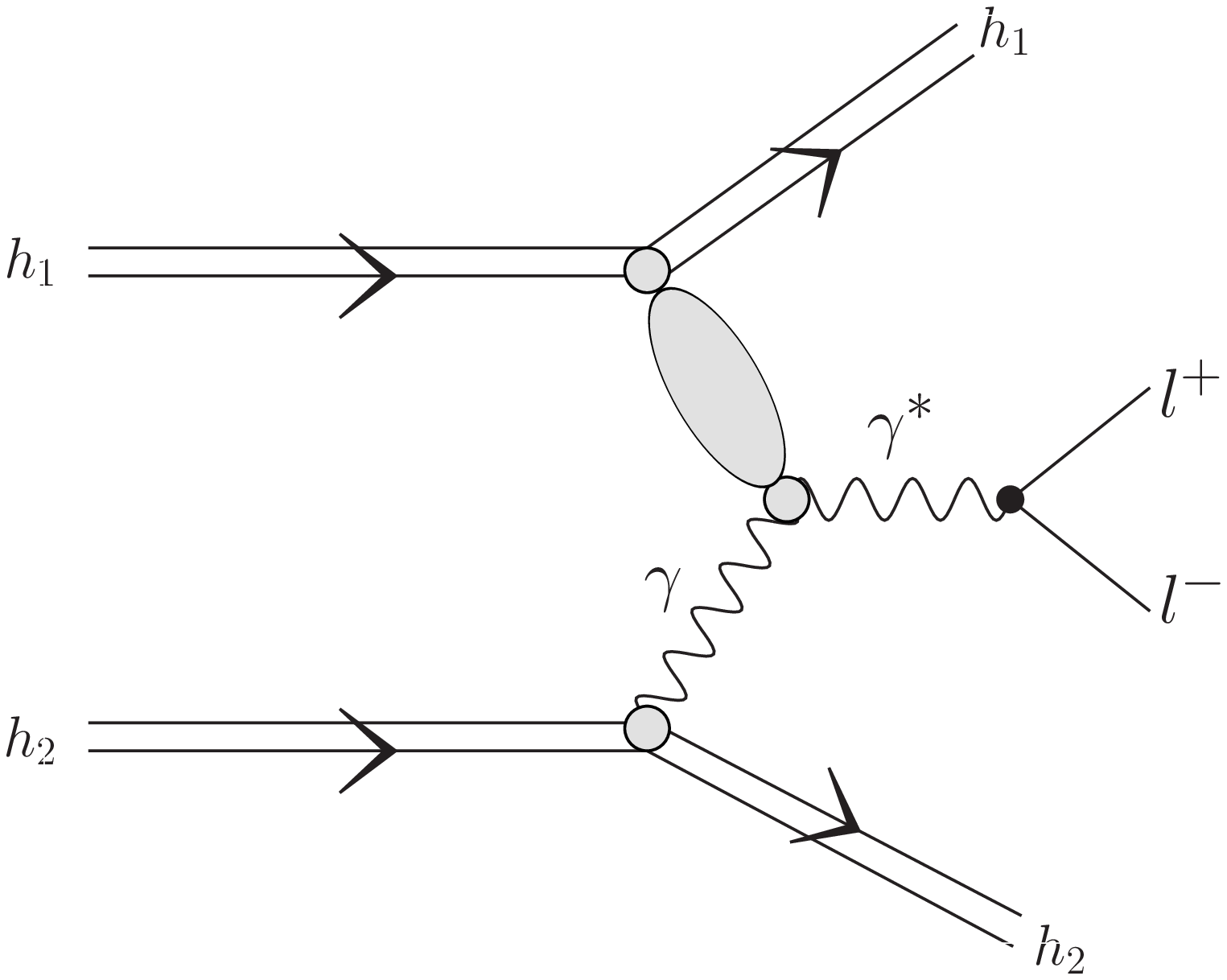}
\hspace{0.5cm}
\includegraphics[width=0.4\textwidth]{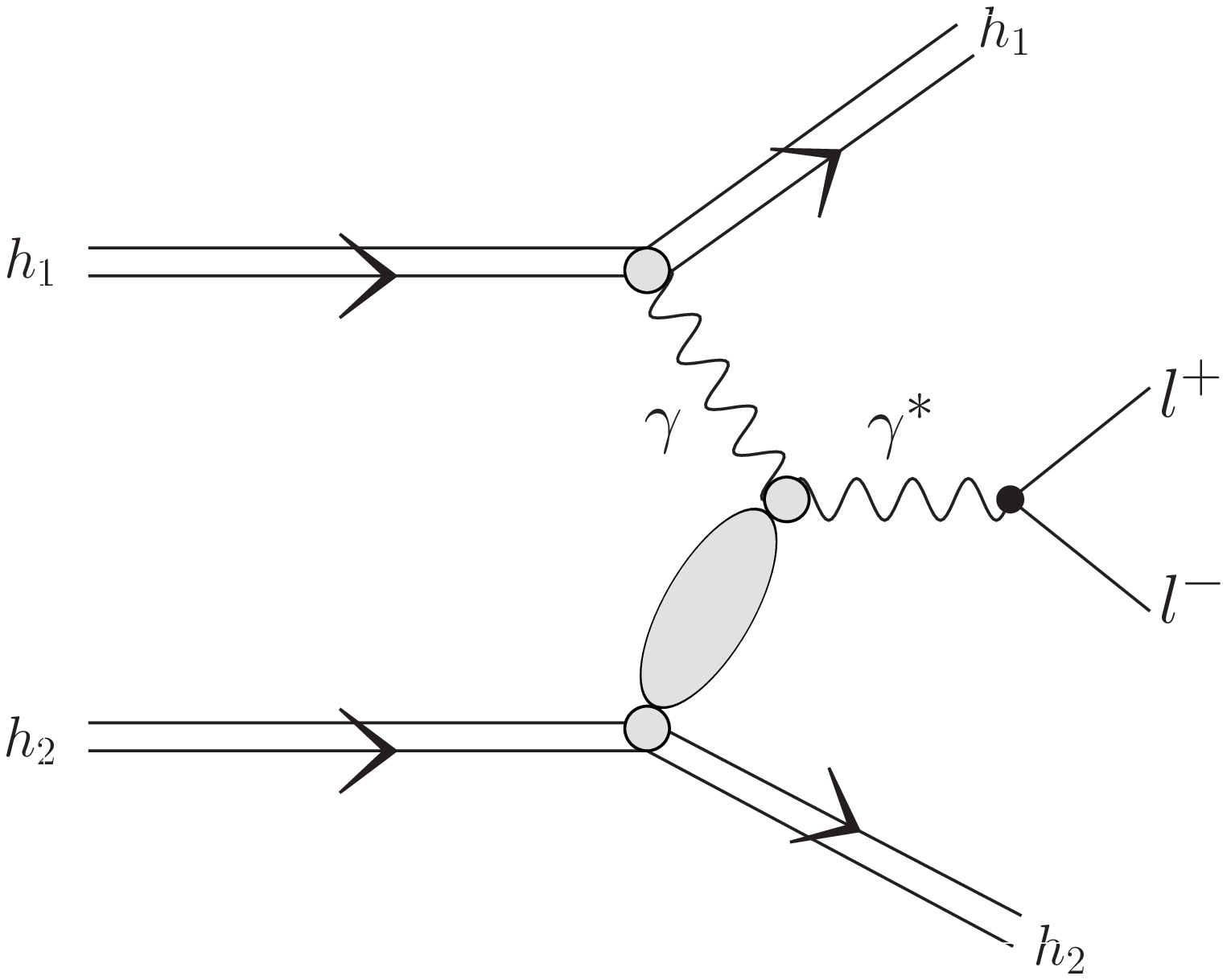}
\end{center}
   \caption{\label{fig:diagrams_pp_to_epempp}
   \small
An example of the non-QED mechanism for the prodution of opposite charge 
leptons in the $p p \to p p l^+ l^-$ reaction. 
}
\end{figure}

As shown in Ref.\cite{SSS10}
the forward $\gamma p \to \gamma^* p$ amplitude is a sum of amplitudes for 
a given flavor and the amplitude for a given flavour contribution can be written as:
\begin{eqnarray}
{\cal M}_f(\gamma p \to \gamma^*(q^2) p)  = W^2\,  4 \pi \alpha_{\mathrm{em}} \, e_{f}^2 \, 2 \, \int_{0}^{1} dz\,\int_{0}^{\infty} \pi dk_{\bot}^2{{\cal A}_{f}(z,k_{\bot}^2,W^2)\over [k_{\bot}^2+m_f^2-z(1-z)q^2-i\varepsilon]}.
\nonumber \\
\end{eqnarray}
The real and imaginary part of the forward Time-like Compton Scattering (TCS)
amplitude takes the form:
\begin{eqnarray}
\Im m  {\cal M}_f(\gamma p \to \gamma^*(q^2) p) &&= W^2\, 16 \pi^2 \alpha_{\mathrm{em}} e_{f}^2
\cdot \Big\{
\theta(4m_f^2 - q^2) \,  \int_{4m_f^2}^{\infty}\, dM^2 {\Im m \,a_{f}(W^2,M^2)\over M^2-q^2}
\nonumber \\
&&+ \theta(q^2 - 4m_f^2) \, 
\Big( \mathrm{PV} 
\int_{4m_f^2}^\infty dM^2 {\Im m\,a_{f}(W^2,M^2) \over M^2 - q^2}
\, + \pi \, \Re e \,a_{f}(W^2,q^2) 
\Big)
\Big\},
\nonumber \\
\Re e  {\cal M}_f(\gamma p \to \gamma^*(q^2) p) &&= W^2\, 16 \pi^2 \alpha_{\mathrm{em}} c_{f}^2
\cdot \Big\{
\theta(4m_f^2 - q^2) \,  \int_{4m_f^2}^{\infty}\, dM^2 {\Re e \,a_{f}(W^2,M^2)\over M^2-q^2}
\nonumber \\
&&+ \theta(q^2 - 4m_f^2) \, 
\Big( \mathrm{PV} 
\int_{4m_f^2}^\infty dM^2 {\Re e\,a_{f}(W^2,M^2) \over M^2 - q^2}
\, - \pi \, \Im m \,a_{f}(W^2,q^2) 
\Big)
\Big\}.
\nonumber \\
\end{eqnarray}
where
\begin{eqnarray}
a_{f}(W^2,M^2) &=& {1 \over M^2} \int_{0}^{{1 \over 4}M^2-m_f^2}\, \frac{dk_{\bot}^2}{J}\,
{\cal A}_{f}(z(M^2,k_{\bot}^2),k_{\bot}^2,W^2) \, .
\end{eqnarray}
Explicit formula for ${\cal A}_{f}$ can be found in \cite{SSS10} 
where we also show its derivation as well as other details. 
The amplitude for the exclusive hadroproduction can be written 
schematicaly as
\begin{eqnarray}
{\cal M}_{p p \to p p l^+ l^-}^{{\lambda}_3 {\lambda}_4}
= e F_1({q_1}^2) ({\bar u_1} {\gamma}^{\mu} u_a)
{ \left(-ig_{\mu \nu} \over t_1 \right)}
{\Sigma}_{{\lambda}_1} {\epsilon}^{\nu}({\lambda}_1)
{\cal M}_{\gamma p \to {\gamma}^{*} p}^{{\lambda}_1 {{\lambda}_1}^{'}}(W_2, t_{2}, M_{ll})
\nonumber \\
{\Sigma}_{{\lambda}_{1}^{'}} {\epsilon}^{{\alpha}^{*}}({{\lambda}_1}^{'})
{\left(-ig_{\alpha \beta} \over s_{34} \right)}
e {\bar u} ({\lambda}_3 , p_3){\gamma}^{\beta} v({\lambda}_4,p_4)
\nonumber
\end{eqnarray}

\begin{eqnarray}
+ e F_1({q_2}^2) ({\bar u_2} {\gamma}^{\mu} u_b)
{ \left(-ig_{\mu \nu} \over t_2 \right)}  
{\Sigma}_{{\lambda}_2} {\epsilon}^{\nu}({\lambda}_2)
{\cal M}_{\gamma p \to {\gamma}^* p}^{{\lambda}_2 {{\lambda}_2}^{'}}(W_1, t_{1}, M_{ll})  
\nonumber \\
{\Sigma}_{{{\lambda}_2}^{'}} {\epsilon}^{{\alpha}^*}({{\lambda}_2}^{'})
{ \left(-ig_{\alpha \beta} \over s_{34} \right)}   
e {\bar u} ({\lambda}_3 , p_3){\gamma}^{\beta} v({\lambda}_4,p_4),
\end{eqnarray}
where $\lambda_{3}, \lambda_{4}$ are helicities of $l^+$ and $l^-$,
respectively. Above $M_{ll}$ is the invariant mass of the lepton pair,
$F_1$ is the Dirac electromagnetic form factor and
${\cal M}_{\gamma p \to {\gamma}^{*} p}^{{\lambda} {\lambda}^{'}}(W, t, M_{ll}) \propto \delta^{\lambda \lambda^{'}} $
are amplitudes for the photon-proton subprocess discussed briefly above.
In the present analysis we omit contributions related to the Pauli
electromagnetic form factors.
Their contribution in the integrated cross section is expected to be negligible. 

Making further manipulations as in \cite{SS_JPsi} we can write the
four-body amplitude for the $\gamma \Pom +  \Pom \gamma$ exchanges in a somewhat 
simplified way
\begin{eqnarray}
{\bf  M}_{ p p \to p p l^+ l^- }^{\lambda_3 \lambda_4}
 \approx \frac{2 e F_1(t_1) {\bf q_1}}{z_1 t_1 \sqrt{1-z_{1}}}
\sum_{\lambda_{1}} {\cal M}_{\gamma p \to \gamma^* p}^{\lambda_{1}}(W_2, M_{ll})
\epsilon_{\mu}^*(\lambda_{1}) 
\exp\left( {B \over 2} t_2 \right)
\frac{e }{M_{ll}^2} 
\bar u(p_3,\lambda_3) \gamma^{\mu} v(p_4,\lambda_4)
\nonumber \\
 +  \frac{2 e F_1(t_2) {\bf q_2}}{z_2 t_2 \sqrt{1-z_{2}}}
\sum_{\lambda_{2}} {\cal M}_{\gamma p \to \gamma^* p}^{\lambda_{2}}(W_1, M_{ll})
\epsilon_{\mu}^*(\lambda_{2})
\exp\left(  {B \over 2} t_1 \right)
\frac{e }{M_{ll}^2}
\bar u(p_3,\lambda_3) \gamma^{\mu} v(p_4,\lambda_4). \nonumber \\
\;
\label{simplified_amplitude}
\end{eqnarray}
Above $z_1$ and $z_2$ are longitudinal momentum fractions
of the intermediate space-like photons with respect
to their parent protons and $\bf q_1$ and $\bf q_2$
are two-dimensional vectors related to the momentum transfer
in the electromagnetic vertiecs.
In the first exploratory calculation presented here we sum only over 
transverse photons, i.e. $\lambda_{1}, \lambda_{2} = \pm 1$.
The $t$-dependence of the amplitude is encoded in the exponential
$ \exp({ B \over 2} t ) $ form factor.
The slope choice of the slope parameter $B$ was discussed in  Ref.\cite{SSS10}.

The cross section is calculated as usually for  a $2 \to 4$ reaction:
\begin{eqnarray}
\sigma =\int \frac{1}{2s} \overline{ |{\cal M}|^2} (2 \pi)^4
\delta^4 (p_a + p_b - p_1 - p_2 - p_3 - p_4)
\frac{d^3 p_1}{(2 \pi)^3 2 E_1}
\frac{d^3 p_2}{(2 \pi)^3 2 E_2}
\frac{d^3 p_3}{(2 \pi)^3 2 E_3}
\frac{d^3 p_4}{(2 \pi)^3 2 E_4}. \nonumber \\
\;
\label{dsigma_for_2to4}
\end{eqnarray}
To calculate the total cross section one has to perform a 
8-dimensional integral numerically.
The details how to conveniently choose kinematical integration variables are explained
in Ref.\cite{LS10}. 

Some distributions initiated by the $\gamma \Pom$ or $\Pom \gamma$ subprocesses
can be calculated with good precision in the equivalent photon approximation (EPA).
A good example of such a distribution is
\begin{eqnarray}
{ d\sigma  \over  dy_{pair}\,d M_{ll}}
= \omega_{1} { d N_{1}(\omega_{1})\over d\omega_{1}} 
{ d \sigma_{\gamma p \to l^{+}l^{-} p} \over dM_{ll}}(W_{2}, M_{ll})
+ \omega_{2} { d N_{2}(\omega_{2})\over d\omega_{2}} 
{ d \sigma_{\gamma p \to l^{+}l^{-} p} \over dM_{ll}}(W_{1}, M_{ll}).
\end{eqnarray}
Above ${dN_{1}\over d\omega_{1}}$ or ${dN_{2}\over d\omega_{2}}$ 
are photon fluxes in the first and second nucleon.
Their explicit form can be found  e.g. in \cite{Drees-Zeppenfeld}.
The differential distributions ${ d\sigma  \over  d M_{ll}}$ were
calculated and shown in Ref. \cite{SSS10}. 
We have found that the EPA distributions are almost identical with 
corresponding ones obtained for the four-body reaction as discussed above.

\subsection{$pp \to p l^+ l^- p$ \,via photon--photon fusion}

\begin{figure}[!ht]    %
\begin{center}
\includegraphics[width=0.4\textwidth]{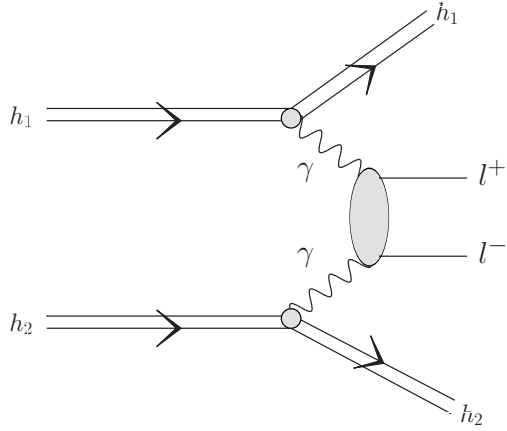}

\end{center}
   \caption{\label{fig:gamma_gamma_diagram}
   \small
The QED $\gamma \gamma$ fusion mechanism of the exclusive lepton pair production. 
}
\end{figure}

Here we present a formalism necessary for the calculation of the
amplitude and cross section for the photon-photon fusion.
The basic mechanism is shown in Fig.\ref{fig:gamma_gamma_diagram}.


The amplitude  for the two-photon $2 \to 4 $ process shown in 
Fig.\ref{fig:gamma_gamma_diagram} can be written as:
\begin{equation}
{\cal M}_{{\lambda}_a {\lambda}_b \to {\lambda}_1 {\lambda}_2 {\lambda}_3 {\lambda}_4}^{p p \to p p l^+ l^- }
={\bar u} (p_1, {\lambda}_1) {\Gamma}_1^{{\mu}_1}(q_1) u(p_a,{\lambda}_a)
\nonumber
\end{equation}

\begin{equation}
{ \left(-ig_{{\mu}_1 {\nu}_1} \over t_1 \right)}
V_{{\lambda}_3 {\lambda}_4}^{{\nu}_1 {\nu}_2}(q_1, q_2, p_3, p_4)
{ \left(-ig_{{\mu}_2 {\nu}_2} \over t_2 \right)}
\end{equation}

\begin{equation}
{\bar u} (p_2, {\lambda}_2) {\Gamma}_2^{{\mu}_2}(q_2) u(p_b,{\lambda}_b),
\nonumber
\end{equation}
where presented below factor\,describes the production amplitude 
of a $l^+ l^-$ pair  with helicities  $\lambda_{3}, \lambda_{4}$ and momenta
$p_{3}, p_{4}$, respectively
\begin{equation}
V_{{\lambda}_3 {\lambda}_4}^{{\nu}_1 {\nu}_2}(q_1, q_2, p_3, p_4) = 
\nonumber \\
e^2 {\bar u} (p_3, {\lambda}_3)
\left[ {\gamma}^{{\nu}_1} { {\hat q_1} - {\hat p_3} - m \over (q_1 - p_3)^2 - m^2} {\gamma}^{{\nu}_2} 
- {\gamma}^{{\nu}_2} { {\hat q_1} - {\hat p_4} + m \over (q_1 - p_4)^2 - m^2}{\gamma}^{{\nu}_1} \right]
v(p_4,{\lambda}_4).
\\
\end{equation}
${\Gamma}_1^{{\mu}_1}(q_1)$ and $ {\Gamma}_2^{{\mu}_2}(q_2)$ are 
vertex functions describing coupling of virtual space like photon to the nucleon,
which can be expressed by the well-known electromagnetic 
Dirac and Pauli form factors of the proton
as:
\begin{eqnarray}
{\Gamma}_1^{{\mu}_1}(q_1) =
{\gamma^{{\mu}_1}} F_1(q_1) + { i {\kappa}_p \over 2 M_p}
{\sigma}^{{\mu}_1 {\nu}_1} q_{{\nu}_1} F_2 (q_1) \; ,
\end{eqnarray}
\begin{eqnarray}
{\Gamma}_2^{{\mu}_2}(q_2) = 
{\gamma^{{\mu}_2}} F_1(q_2) + { i {\kappa}_p \over 2 M_p}
{\sigma}^{{\mu}_2 {\nu}_2} q_{{\nu}_2} F_2 (q_2) \; ,
\end{eqnarray}
where
\begin{eqnarray}
{\sigma}^{\mu \nu}= { i \over 2} ({\gamma}^{\mu} {\gamma}^{\nu} - {\gamma}^{\nu} {\gamma}^{\mu}),
\nonumber \\
q_1 {\equiv} p_1 - p_a,
\nonumber \\
q_2 {\equiv} p_2 - p_b.
\end{eqnarray}

Using the Gordon decomposition one can simplify the tensorial structure
\begin{eqnarray}
{\Gamma}_1^{{\mu}_1}(q_1) = 
(F_1(q_1) + {\kappa}  F_2 (q_2)) {\gamma}^{{\mu}_1}
- { {\kappa} F_2 (q_1)\over 2 m_N} (p_a + p_1)^{{\mu}_1} \; ,
\end{eqnarray}
\begin{eqnarray}
{\Gamma}_2^{{\mu}_2}(q_2) = 
(F_1(q_2) + {\kappa}  F_2 (q_2)) {\gamma}^{{\mu}_2}
- { {\kappa} F_2 (q_2)\over 2 m_N} (p_b + p_2)^{{\mu}_2} \; .
\end{eqnarray}
As for the diffractive case in the present paper we neglect contributions related to 
the Pauli form factors which are very small for the integrated cross section.

\subsection{Results}

\begin{figure}[!ht]    %
\begin{center}
\includegraphics[width=0.5\textwidth]{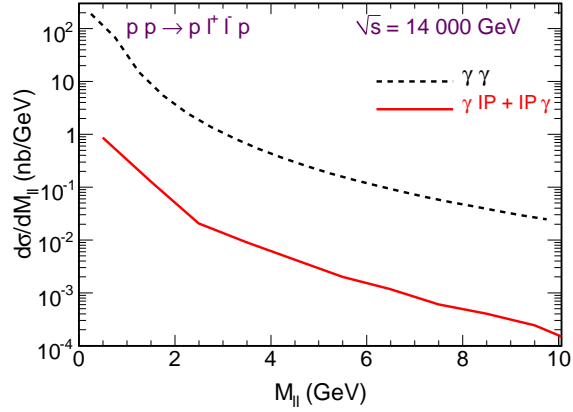}
\end{center}
   \caption{\label{fig:m_ll}
   \small
Dependence of the cross section on the dilepton invariant mass for the 
$\gamma \gamma$ (dashed line) and for the diffrative mechanism (solid line).
}
\end{figure}

In this section we shall present results for exclusive
diffractive mechanism discussed above.
We shall show differential cross sections 
for $\mu^{+} \mu^{-}$ production via $\gamma \Pom$ or $\Pom \gamma$ 
exchange and for comparison via $\gamma \gamma$ fusion 
\footnote{In contrast to inclusive diffractive DY the cross section for exclusive
production of $e^{+} e^{-}$ pairs is significantly bigger than that for
$\mu^{+} \mu^{-}$ production.}.
Here we shall concentrate on the LHC energy  $\sqrt{s}$= 14 TeV.
Let us start from the dilepton invariant mass distribution
shown in Fig.\ref{fig:m_ll}. 
The diffractive contribution is about two-orders of magnitude
smaller that that for the photon-photon fusion.
The shape of the distribution is rather similar.
We do not include here absorption effects neither for the $\gamma \gamma$
nor for the $\gamma \Pom$ ($\Pom \gamma$).
In both cases they are rather small (see e.g. \cite{SS_JPsi}).

As for the inclusive case we also show distribution for
lepton pair rapidity. The diffractive component is in addition decomposed
into separate contributions corresponding to $\gamma \Pom$ fusion
( right bump) and $\Pom \gamma$ fusion (left bump).
It can be shown that without absorption effects the two contributions
add incoherently in the lepton pair rapidity \footnote{This is not true
for other distributions, in particular for azimuthal angle correlations
between ougoing protons (see e.g.\cite{SS_JPsi}).}.


\begin{figure}[!ht]    %
\begin{center}
\includegraphics[width=0.45\textwidth]{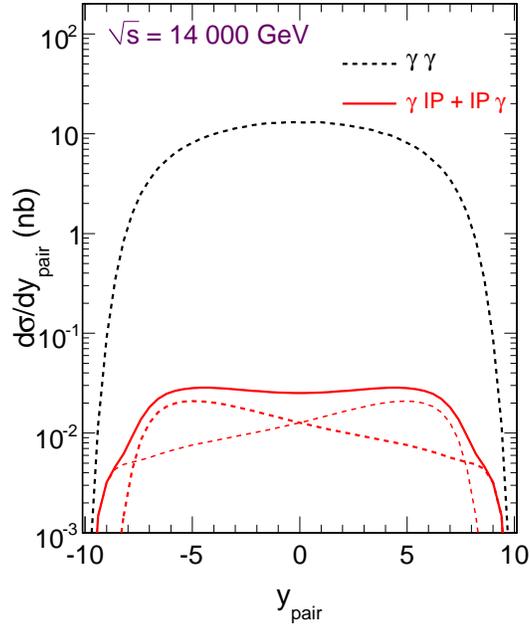}
\end{center}
   \caption{\label{fig:y_pair}
   \small
Lepton pair rapidity distribution for the $\gamma \gamma$ (dashed line) and for
the diffractive mechanism (solid line) which is further decomposed into 
$\gamma \Pom$ and $ \Pom \gamma$ contributions. 
}
\end{figure}

\begin{figure}[!ht]    %
\begin{center}
\includegraphics[width=0.5\textwidth]{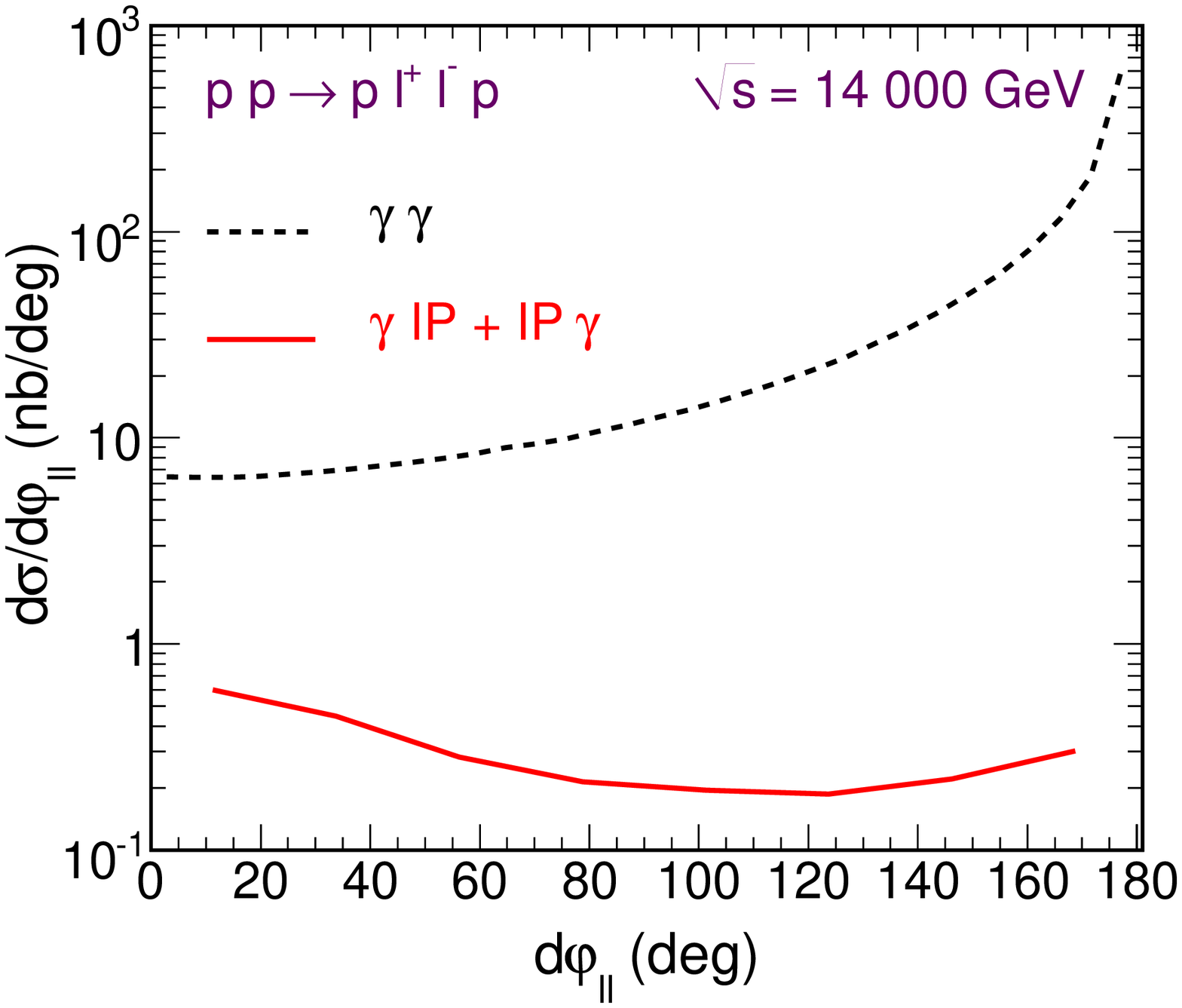}
\end{center}
   \caption{\label{fig:phi_ll}
   \small
Distribution in relative azimuthal angle between outgoing leptons.
}
\end{figure}

In Fig.\ref{fig:phi_ll} we present azimuthal correlations 
between the outgoing leptons for the
$\gamma \gamma$ fusion (dashed line) and for the $\gamma \Pom + \Pom \gamma$ 
exchanges (solid line).
Azimuthal angle distribution for the $\gamma \gamma$ process peaks sharply at 
$\phi \sim$ 180 $^o$ but for the $\gamma \Pom + \Pom \gamma$ process 
leptons prefer to go into the same hemisphere.
This distribution could be therefore used for imposing cuts in order
to enhance the contribution of the new diffractive photoproduction
mechanism.


\begin{figure}[!ht]    %
\begin{center}
\includegraphics[width=0.5\textwidth]{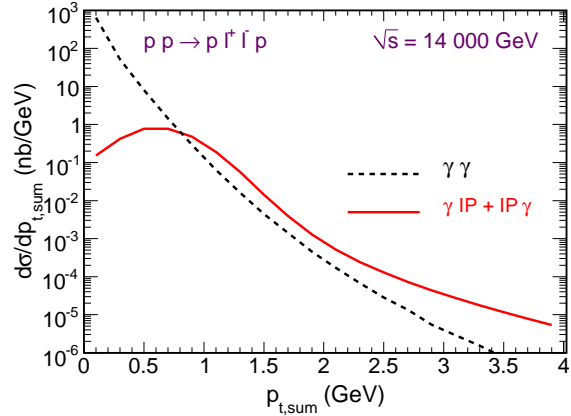}
\end{center}
   \caption{\label{fig:pt_sum}
   \small
Dependence on the transverse momentum of the dilepton pair ($p_{t,sum}$) for
the diffractive (solid line) and photon-photon (dashed line) contributions.
}
\end{figure}

In all the distributions presented above the $\gamma \gamma$
mechanism dominates over the diffractive one.
Can the diffractive mechanism be identified experimentally?
The leptons in $\gamma \gamma$ process are emitted preferentially
back-to-back and their transverse momenta almost cancel.
This means that for this process transverse momentum
of the pair should be small. It is not necessarily so for the
diffractive mechanism where the transverse momentum kick to a proton due to the pomeron
(gluonic ladder) exchange is much bigger than that due to photon exchange.
In Fig.\ref{fig:pt_sum} we show distribution in transverse momentum of
the dilepton pair
($\overrightarrow{p_{t,}}_{sum} = \overrightarrow{p_{1t}} + \overrightarrow{p_{2t}}$).
As expected the photon-photon contribution
dominates at small transverse momenta of the pair, while the photon-pomeron
(pomeron-photon) contributions at transverse momenta larger than about 1
GeV.
We therefore think that imposing a cut on the variable would be useful and perhaps
 necessary to identify the diffractive photoproduction mechanism.

\begin{figure}[!ht]    %
\begin{center}
\includegraphics[width=0.4\textwidth]{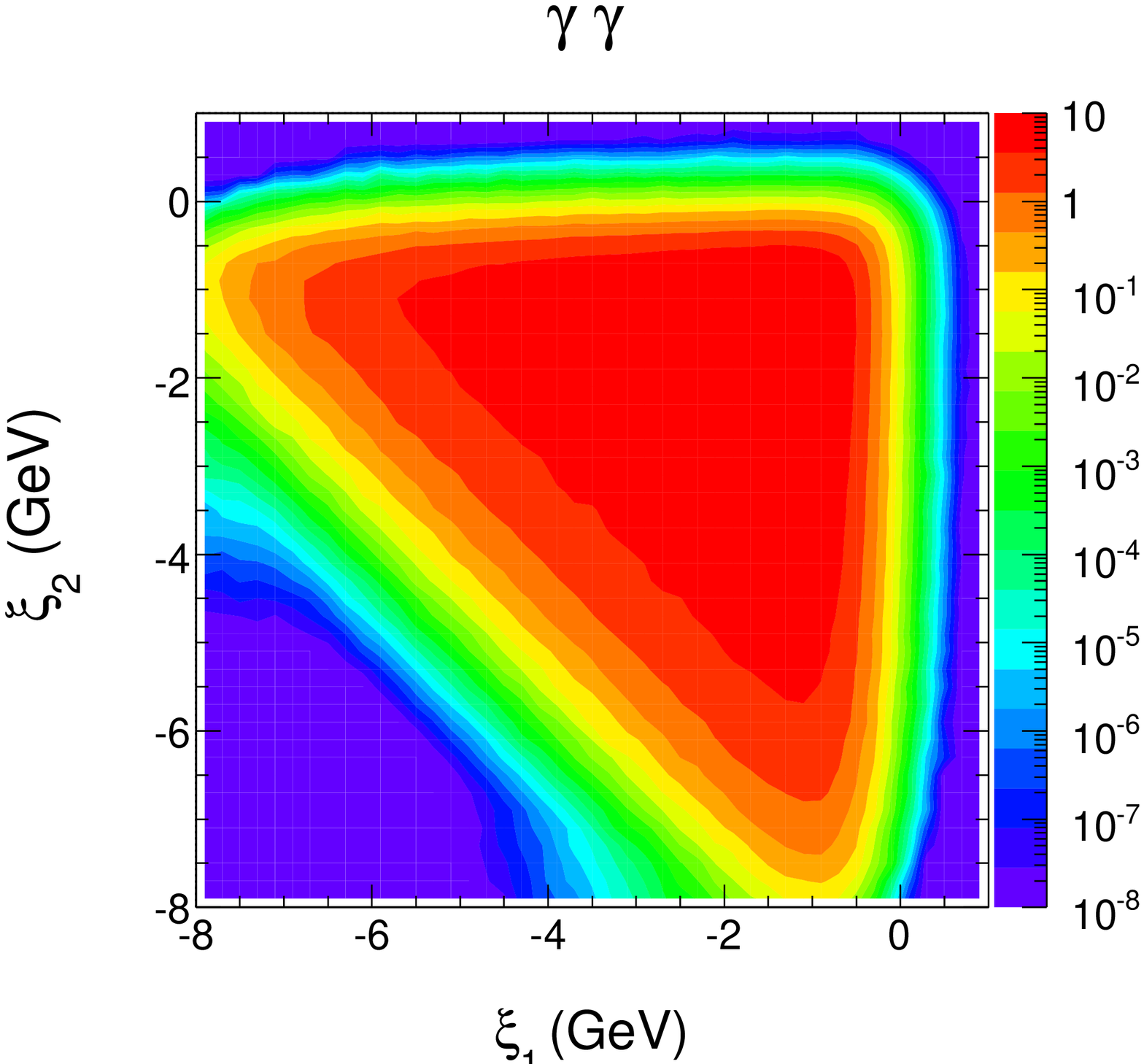}
\includegraphics[width=0.4\textwidth]{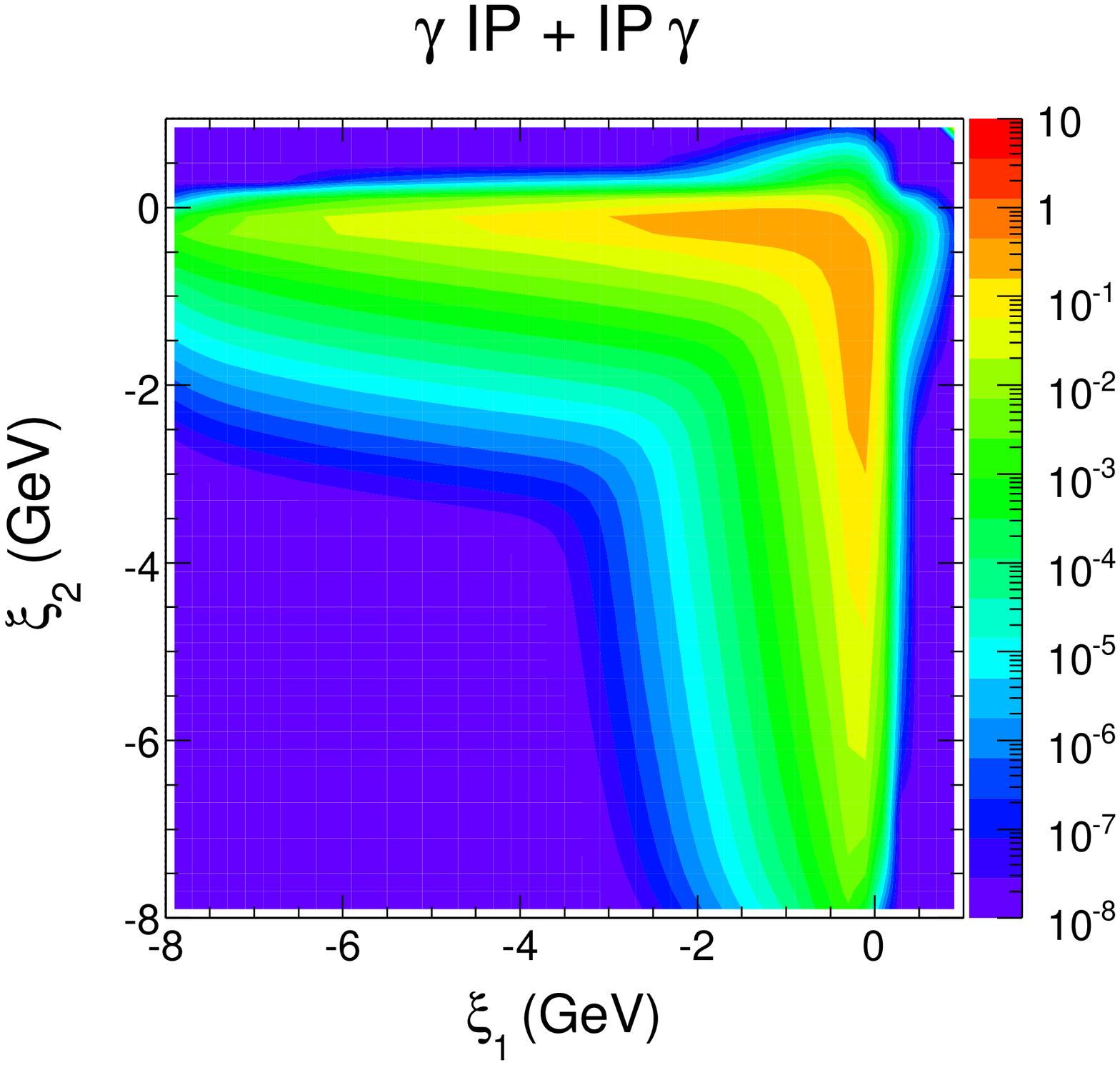}
\end{center}
   \caption{\label{fig:xi1_xi2_gg_gp}
   \small
Transverse momentum correlations of outgoing protons
for the $\gamma \gamma$ fusion (left panel) and  for the 
$\gamma \Pom + \Pom \gamma$  exchange (right panel).
}
\end{figure}

Can we further ``pin down'' the photoproduction mechanism?
Let us consider now two-dimensional correlations for outgoing
particles.
Let us start with correlations between transverse momenta of outgoing protons.
Since transverse momenta of outgoing protons are rather small (photon or
pomeron exchange) we shall use 
$\xi_{1} = \log_{10}[$p$_{1t}$/1 GeV] and $\xi_{2} = \log_{10}[$p$_{1t}$/1 GeV]
instead of transverse momenta.
In Fig.\ref{fig:xi1_xi2_gg_gp} we show two-dimensional correlations
in the ($\xi_{1},\xi_{2}$) space. Different pattern can be seen for the
$\gamma \gamma$ and diffractive mechanisms. It is not clear to us at present
if the measurement of transverse momenta of protons will be precise
enough to impose cuts in the two-dimensional space.


\begin{figure}[!ht]    %
\begin{center}
\includegraphics[width=0.4\textwidth]{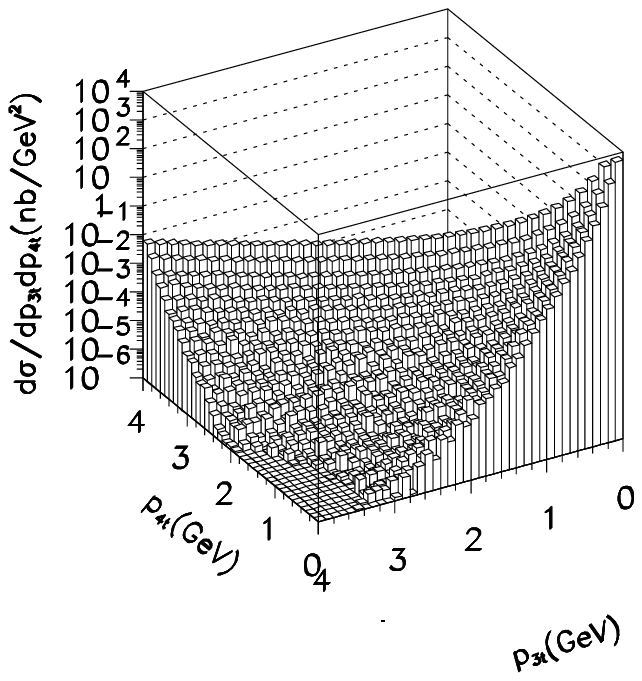}
\includegraphics[width=0.4\textwidth]{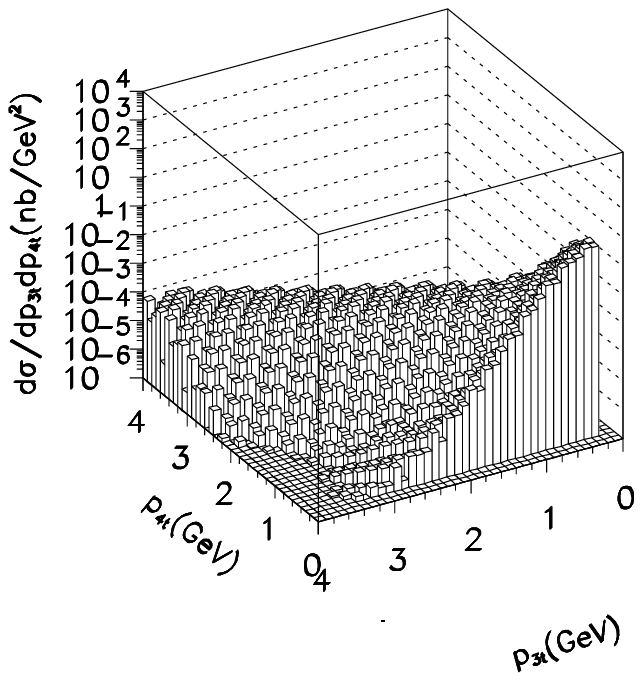}
\end{center}
   \caption{\label{fig:pt3_pt4_gg_gp}
   \small
Correlations in transverse momenta of outgoing leptons
for the $\gamma \gamma$ fusion (left panel) and  for the 
$\gamma \Pom + \Pom \gamma$ exchange (right panel).
}
\end{figure}

We do similar analysis for transverse momenta of outgoing muons.
Fig.\ref{fig:pt3_pt4_gg_gp} shows two-dimensional distribution in transverse
momentum of outgoing leptons. A strong correlation between transverse
momentum of the negative and positive muon can be seen.
The further from the diagonal the bigger fractional contribution
of the diffractive mechanism.
We conclude that this figure contains essentially similar information as 
the distribution in transverse momentum of the dilepton pair shown in 
Fig.\ref{fig:pt_sum}.


\begin{figure}[!ht]    %
\begin{center}
\includegraphics[width=0.4\textwidth]{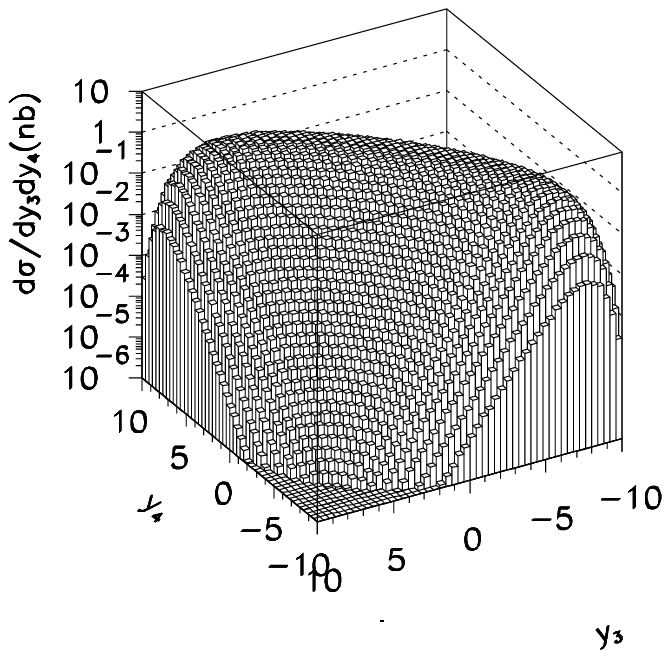}
\includegraphics[width=0.4\textwidth]{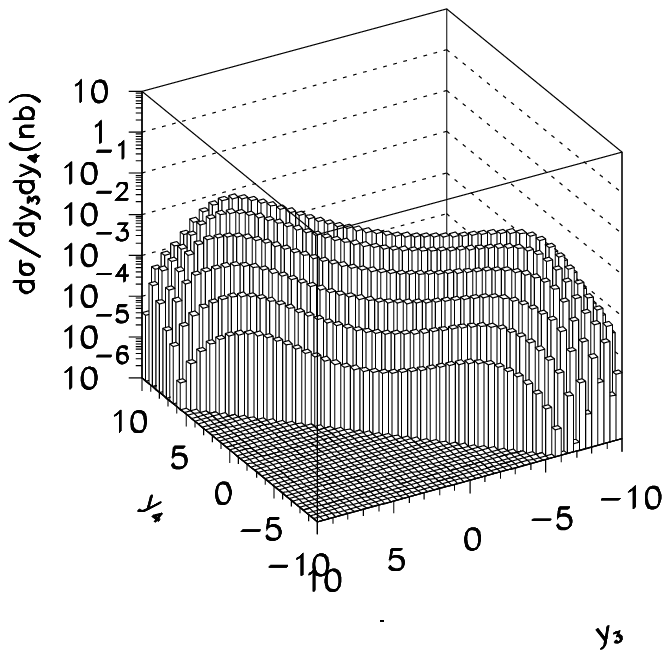}
\end{center}
   \caption{\label{fig:y3_y4_gg_gp}
   \small
Rapidity correlations of leptons for the
$\gamma \gamma$ fusion (left panel) and for the $\gamma \Pom + \Pom \gamma$ 
exchange (right panel).
}
\end{figure}

In analogy to the inclusive case we consider correlations
in the lepton rapidity space.
The situation on the two-dimensional plane $(y_3, y_4)$ is shown
in Fig.\ref{fig:y3_y4_gg_gp} for the 
$\gamma \gamma$ fusion (left panel) and for the $\gamma \Pom + \Pom \gamma$ 
exchange (right panel).
We observe that the correlations for the diffractive mechanism
are stronger than that for the $\gamma \gamma$ fusion.
Therefore one could impose further cut
on the difference of the lepton rapidities: $y_{diff} \equiv y_3 - y_4$.


\begin{figure}[!ht]    %
\begin{center}
\includegraphics[width=0.6\textwidth]{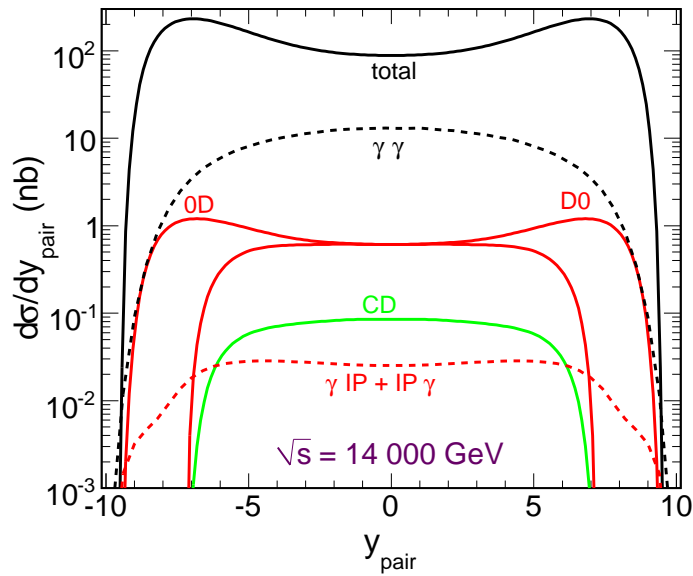}
\end{center}
   \caption{\label{fig:exclusive_vs_inclusive}
   \small
Distribution in lepton pair rapidity for all processes considered in 
the present paper at the nominal LHC energy $\sqrt{s}$= 14 000 GeV. The inclusive diffractive
processes are shown by the solid lines and the exclusive ones by the dashed lines.
Here we have included gap survival factors as explained in the text.
}
\end{figure}

How big are cross sections for the exclusive mechanisms discussed
in this section in comparison to those for inclusive cross sections
calculated in the Ingelman-Schlein model corrected for absorption.
In Fig.\ref{fig:exclusive_vs_inclusive} we have collected lepton pair rapidity 
distributions for different processes:
ordinary nondiffractive Drell-Yan, single and central diffractive Drell-Yan,
exclusive production of dilepton pair via the $\gamma \gamma$ fusion and 
via $\gamma \Pom + \Pom \gamma$ exchange.
In this plot we include absorption effects discussed in the previous section.
We observe that the cross section for the $\gamma \gamma$ mechanism is
larger than that for the single and central diffractive ones.
On the other hand, the cross section for exclusive diffractive production
is only slightly smaller than that for the central diffractive mechanism.

Further studies are clrearly needed in order to demonstrate whether measurements of
the cross section of the discussed mechanisms are possible.
 

\section{Conclusions}

We have calculated distributions in lepton rapidity, lepton
transverse momentum as well as dilepton invariant mass
for inclusive single and central diffractive production
of dileptons in proton-proton collisions. 
In this calculations we have used diffractive parton distributions
found from the analysis of the proton diffractive structure function
and dijet production in deep inelastic scattering. 
The distributions have been compared with the corresponding
distributions for ordinary nondiffractive Drell-Yan process.
The distribution in rapidity for the single-diffractive process
is very similar to that for the nondiffractive case.
The single diffractive mechanism constitutes about a  percent 
of the inclusive mechanism. The cross section for central diffractive 
mechanism is smaller than that for single diffractive one by one order of magnitude.

In our approach the ratio of the diffractive to the total cross section
for the dilepton production only slightly depends on the center-of-mass
energy and the dilepton mass. This is in evident contrast to
earlier predictions made within the dipole approach.
Experimental studies would clearly shed more light on the issue
and would help in understanding the diffractive mechanism in hadronic processes,
 certainly not fully understood so far.

We have also calculated several differential distributions
for exclusive diffractive production of dileptons. Here the
photon-pomeron (pomeron-photon) is the driving mechanism.
We have applied here a formalism used previously for the
$\gamma p \to l^+ l^- p$ reaction.

This formalism was 
previously successfully tested for exclusive production of vector mesons.
The distributions for the diffractive exclusive process were
compared with corresponding distributions for the QED photon-photon
mechanism. We have found regions of the phase space where the
diffractive mechanism dominates over the QED one.
Several differential distributions have been shown and discussed.
Experimental identification of the exclusive diffractive process is very
important in the context of the proposal to use the QED photon-photon fusion
to monitor luminosity at the LHC.
Clearly further Monte Carlo studies are necessary.

\vspace{1cm}

{\bf Acknowledgments}

This work was partially supported by the
MNiSW grants: N202 2492235 and N N202 236937. 
A discussion with Wolfgang Sch\"afer  and his  collaboration on the dilepton production in
the $\gamma p \to l^+ l^- p$ reaction as well as a discussion with 
Poman Pasechnik is kindly acknowledged.


\end{document}